\documentclass[preprint,11pt,authoryear]{elsarticle}

\usepackage{natbib}
\usepackage{lineno}
\usepackage{listings}
\usepackage{float}

\usepackage{amssymb}
\usepackage{amsmath}
\usepackage{hyperref}
\usepackage{cleveref}
\usepackage{eurosym}
\usepackage[english]{babel}

\usepackage{amsfonts}
\usepackage{amssymb}
\emergencystretch=\maxdimen
\nolinenumbers
 
\journal{Journal of South American Earth Science}

\begin{document}

\begin{frontmatter}

\title{Simulated Annealing for Volcano Muography}

\author[adir,bdir,cdir]{A. Vesga-Ram\'{\i}rez\corref{mycorrespondingauthor}}
\cortext[mycorrespondingauthor]{Corresponding author}
\ead{alejandravesga@cnea.gov.ar}
\author[cdir]{J.D. Sanabria-G\'omez}
\author[ddir]{D. Sierra-Porta}
\author[edir]{L. Arana-Salinas}
\author[bdir,fdir,gdir,hdir]{H. Asorey}
\author[idir]{V. A. Kudryavtsev}
\author[bdir,fdir,jdir]{R. Calder\'on-Ardila}
\author[cdir,kdir]{L.A. N\'u\~nez}

\address[adir]{International Center for Earth Sciences, Comisi\'on Nacional de Energ\'{\i}a At\'omica, Buenos Aires, Argentina;}
\address[bdir]{Consejo Nacional de Investigaciones Cient\'{\i}ficas y T\'ecnicas, Argentina;}

\address[cdir]{Escuela de F\'{\i}sica, Universidad Industrial de Santander, Bucaramanga, Colombia;}

\address[ddir]{Departamento de F\'isica, Universidad de Los Andes, Bogot\'a, Colombia;}

\address[edir]{Academia de Protecci\'on Civil y Gesti\'on de Riesgo, Universidad Aut\'onoma de la Ciudad de M\'exico, Colegio de Ciencias y Humanidades, Ciudad de M\'exico, M\'exico;}

\address[fdir]{Instituto de Tecnolog\'{\i}as en Detecci\'on y Astropart\'{\i}culas, Centro At\'omico Constituyentes, Buenos Aires, Argentina;}

\address[gdir]{Centro At\'omico Bariloche, Comisi\'on Nacional de Energ\'{\i}a At\'omica, Bariloche, Argentina;}

\address[hdir]{Departamento F\'isica M\'edica, Comisi\'on Nacional de Energ\'ia At\'omica, Argentina;}

\address[idir]{Department of Physics and Astronomy, University of Sheffield, Sheffield, S3 7RH, United Kingdom;}

\address[jdir]{Instituto SABATO, Universidad Nacional de San Mart\'in, Buenos Aires, Argentina.}

\address[kdir]{Departamento de F\'isica, Universidad de Los Andes, M\'erida, Venezuela;}

\begin{abstract}
Muography or muon radiography is a non-invasive emerging image technology relying on high energy atmospheric muons, which complements other standard geophysical tools to understand the Earth’s subsurface.

This work discusses a geophysical inversion methodology for volcanic muography, based on the Simulated Annealing algorithm, using a semi-empirical model of the muon flux to reach the volcano topography and a framework for the energy loss of muons in rock.

The Metropolis-Simulated-Annealing algorithm starts from an ``observed'' muon flux and obtains the best associated inner density distribution function inside inside a syntetic model of the Cerro Mach\'in Volcano (Tolima-Colombia).
The estimated initial density model was obtained with GEOMODELER, adapted to the volcano topography.  We improved this model by including rock densities from samples taken from the crater, the dome and the areas associated with fumaroles.

 In this paper we determined the minimum muon energy (a function of the arrival direction) needed to cross the volcanic building, the emerging integrated flux of muons, and the density profile inside a model of Cerro Machin. The present inversion correctly reconstructed the density differences inside the Mach\'in, within a $1\%$ error concerning our initial simulation model, giving a remarkable density contrast between the volcanic duct, the encasing rock and the fumaroles area.
\end{abstract}




\end{frontmatter}

\section{Introduction}
\label{sec:intro}
A technique known as muon radiography, muon tomography --or merely muography-- is emerging with many applications ranging from geosciences to nuclear safety, civil engineering and archaeology (see  \cite{PROCUREUR2018, kaiser2019muography, BONECHI2020100038, ThompsonEtal2020}, and references therein). This technique is based on measuring the attenuation of the atmospheric muon flux travelling through the material  \citep{Tanaka2007b, lesparre2010geophysical, okubo2012imaging}. Atmospheric muons originate from the decay of charged pions, kaons, and other mesons through weak interactions processes, produced while cosmic rays cross Earth's atmosphere. The small cross-section of muon interaction with matter  \citep{barrett1952interpretation} due to its mass being $200$ times higher than that of the electron, and the muon energy spectrum --extending to high energies \citep{gaisser1990cosmic}-- results in the high penetrating power of muons and their ability to cross hundreds of meters of rock (although suffering a significant flux attenuation). There are two main mougraphy techniques: {\it absorption-based} or {\it transmission muography}  --based on the opacity of the material along that line of sight-- and {\it scattering-based muography} --which measures the average muon deflection angle ($\Delta \theta$) due to Coulomb scattering of the muons with the nucleus of the scanning materials-- \citep{PROCUREUR2018, BONECHI2020100038}.

As pointed by R. Kaiser \citep{kaiser2019muography}, the muography community is very active. It is transiting the path from a research field to a technological innovation area, where half a dozen companies are providing services and products in several commercial niches. In Latin America, three initiatives from Mexico, Colombia and Argentina are starting to cooperate in several research projects.  

A quick search\footnote{Searching: muography OR "muon tomography" OR "muon radiography"} in google scholar found almost 1,200 publications\footnote{1,177 publications including patents distributed as 221 in 2016; 226 in 2017; 235 in 2018; 264 in 2019 and 231 in 2020} with 194 US patents\footnote{33 in 2016; 34 in 2017; 32 in 2018; 42 in 2019 and 53 in 2020} in the last five years.  The cost of this competitive technology depends on both, the technique (scattering-based muography or absorption-based muography) and the required resolution ($> 10$~mrad to $< 10$~mrad), ranging from $\approx 10 K$\euro/m$^2$ to $\approx 50 K$\euro/m$^2$ (see Table 1 in reference \cite{BONECHI2020100038}). 

Volcanoes are structures of great interest from both geological and geophysical points of view. Seismic, gravimetric and other standard geophysical methods have been applied to model inner volcanic structures \citep{mcnutt1996seismic}, while geological, geochemical and geophysical studies are routinely made to understand their composition and past behaviour. Rock drilling is one of the best ways of collecting information but is expensive and limited to the area and depth of drilling. 

Active volcanoes are those which have erupted in the recent past (last 10000 years) and could represent a hazard for their surrounding populations. The knowledge of the internal structure of active volcanoes is crucial in assessing their potential impact. However, this remains one of the most challenging geophysics problem because defining the structure of a highly composite volcano requires dense data sampling in severe field conditions.  

In Colombia, it is worth mentioning the 3D model of the Nevado del Ruiz volcano which integrates geological, geochemical and geophysical (seismic and magneto-telluric) information  \citep{gonzalez2015nevado}. However, several authors \cite {munoz1992local, lesparre2012density} have pointed out certain limitations in spatial resolution of geophysical methods, related for example to the wavelength in seismic surveys.

In vulcanology, muon radiography measures the atmospheric muon flux attenuated by rock volumes of different densities, projecting images of volcanic conduits at the top of the volcanic structure, which are critical in understanding possible eruption dynamics. Obtaining the density profile in volcanoes requires comparing the muon flux detected by a properly calibrated instrument with the flux expected and validated by a detailed simulation. When the detected flux is higher than expected, the average density along that path must be less than the one initially considered and vice versa  \citep{Tanaka2007b, lesparre2010geophysical, okubo2012imaging, calderon2020study}. 

More than a dozen active volcanoes in Colombia, representing significant risks to the nearby population\, \citep{Cortes2016, Agudelo2016, Munoz2017}, have motivated local research groups to explore possible applications of muography to inland geological structures. Colombian inland volcanoes are commonly surrounded by other geological structures that screen the atmospheric muon flux, and only a few potential observation points are available. Some ideas had been reported around the possibility of designing a muon telescope to study the Galera volcano in the southern part of Colombia \citep{TapiaEtal2016, GuerreroEtal2019}. In Bogotá, at the centre of the country, J.S. Useche-Parra and C. A. Avila-Bernal designed and built a hodoscope prototype to measure muon flux crossing Monserrate Hill. Through the interpolation of stopping power data, they determine the muon incident energy's attenuation function and estimate the muon flux as a function of the mountain's location \citep{ParraAvila2019}. The Astroparticle research group at the Universidad Industrial de Santander --in cooperation of several institutions in Argentina--, developed a research muography program, designing and constructing instrumentation to study the Colombia Volcanoes, and civil structures \citep{AsoreyEtal2017, SierraportaEtal2018, PenarodriguezEtal2018B, PenarodriguezEtal2019, VasquezEtal2020, PenarodriguezEtal2020}. The landscape in Colombia --and indeed in all other Andean volcanoes-- is very different. Most of the country active volcanoes are along the Cordillera Central, surrounded by higher altitudes shielding cosmic ray flux. Thus, we developed a methodology to identify the most feasible volcano candidates, identifying the best candidates and the possible observation point to study with this technique \citep{VesgaramirezEtal2020}.
  
The present work describes an optimization method for obtaining the best density distribution inside the volcano by measuring the muon flux emerging from the geological structure. Some previous work has considered this inverse problem using a Bayesian framework \citep{barnoud2019bayesian, nishiyama20173d, nishiyama2014integrated} or traditional local optimization methods \citep{lelievre2019joint, rosas2017three, davis2011joint}. In this study, we use a global optimization inversion method based on Simulated Annealing Algorithm (SAA) --applying the Metropolis criterion-- to obtain the global optimum for the density distribution function inside the volcano. The name of this algorithm, inspired by annealing in metallurgy, was proposed in 1983 by S. Kirkpatrick, C.D. Gelatt Jr. and M. P. Vecchi when they solved the travelling salesman problem applying this optimization scheme\citep{kirkpatrick1983optimization}. The SAA inversion has some advantages over other approach based on local inversion methods. It is computationally efficient, can solve linear and no linear problems, underdetermined inverse problems, can more easily implement complex a priori information, and does not introduce smoothing effects in the final density structure model \citep{nagihara2001three}. 

We have organized the paper as follows. In the next section, we briefly describe our hybrid muon telescope and the criteria used to select the muography site in Colombia. Section \ref{GeologicalForwardModel} develops two essential inputs to the geophysical inversion: the geological model of the Cerro Mach\'in Volcano and the initial density distribution emerging from the forward modelling. Section \ref{InverseModelling} implements the geophysical inversion by using the SSA, while Section \ref{ImplementingInversion} discusses the results obtained. Finally, in section \ref{Conclusions}, we outline some final remarks and conclusions along with possible future work. 
\section{Muography in Colombia}
\label{MuographyColombia}
Colombia has more than a dozen active volcanoes along the Cordillera Central, the highest of the three branches of the Colombian Andes.  Most of these volcanoes represent a significant risk to the nearby population in towns and/or villages\citep{Cortes2016,Agudelo2016,Munoz2017} and have caused major disasters.

In this section, we briefly describe our hybrid \textsl{Mu}on \textsl{Te}lescope, MuTe, and the selected muography observation site at Cerro Mach\'in in the department of Tolima. Details of the instrument operation and calibration are considered elsewhere \citep{AsoreyEtal2017, PenarodriguezEtal2019, VasquezEtal2020, PenarodriguezEtal2020}, while the reasons in choosing Cerro Mach\'in among 13 other Colombian volcanoes can be found in reference \citep{VesgaramirezEtal2020}.

\begin{figure}[h!]
\begin{center}
	\includegraphics[scale=0.30]{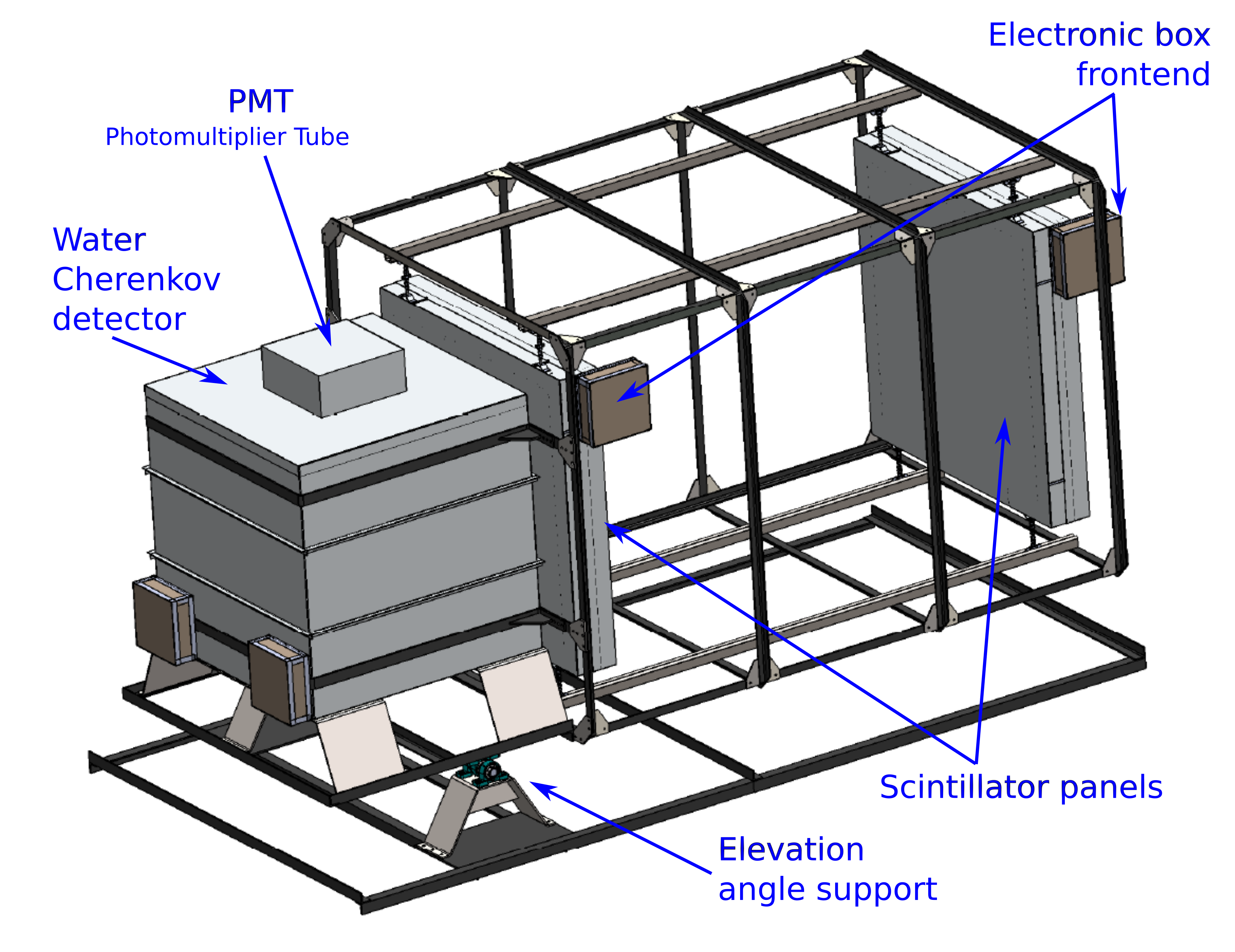} 
		  \caption[MuTe hybrid telescope] {Schematic of MuTe: a hybrid telescope with two scintillator array panels --devised to identify muon trajectories-- and a water Cherenkov detector which filters most of the backwards \& background noise of muography.}
  \label{detector}
\end{center}   
\end{figure}

\subsection{MuTe: a hybrid Muon Telescope}
\label{sec:scint}
There are three main types of detectors implemented for volcano muography: nuclear emulsion, scintillation and gas detectors. Each one has its \textit{pros} and \textit{cons} as described in references \citep{TanakaOlah2019, Tanaka2016}. 

MuTe is a hybrid telescope with two detectors combined:
\begin{itemize}
    \item \textbf{Two-panel-hodoscope:} Hodoscopes are the most common detectors designed and implemented for volcano muography. They consist of two panels devised to identify muon trajectories. Inspired by the experiences of other volcano muography experiments  \citep{UchidaTanakaTanaka2009,GibertEtal2010}, we have designed two X-Y  arrays of $30~\times~30$ plastic scintillating strips ($120$~cm$~\times~4$~cm$~\times~1$~cm). Each array  ($4$~cm$~\times~4$~cm~=~$16$~cm$^2$) has $900$~pixels, giving a surface detection area of $14400$~cm$^2$. The two panels can be separated up to $D=250$~cm  \citep{VasquezEtal2020}. 
    \item \textbf{Water Cherenkov Detector:} The Water Cherenkov Detector, WCD, indirectly detects charged particles by the Cherenkov photons generated by relativistic particles traveling through water.  The MuTe WCD is a purified water cube of $120$~cm side, located behind the rear scintillator panel (see Figure \ref{detector}), which filters most of the backwards and the background noise of muography, discriminating the muonic from the electromagnetic component of the atmospheric showers produced by cosmic rays \citep{VasquezEtal2020}. 
\end{itemize}

\subsection {The MuTe Selection Sites}
In Colombia, various active volcanoes are found along the Cordillera Central, with neighbouring geological structures of higher altitude. This complex surrounding topographic environment obstructs the tilted cosmic ray fluxes increasing the overall distance the muons travel through rock, and distorting the estimated density distribution inside the volcano.  Thus, we used the topographical surface map of the Mach\'{\i}n volcano\footnote{From Alaska Satellite Facility Vertex \url{https://vertex.daac.asf.alaska.edu/}} with a resolution of $12.5$~m$~\times~12.5$~m, and with a ray-tracing algorithm calculated all the possible distances associated with muon paths crossing only the dome. Then, we determined the muon energy losses to estimate muon flux at the detector. 

\begin{table}[ht]
\centering
	\caption{Observation points at Cerro Mach\'{\i}n volcano. Our  modelling will be implemented simulating the muon flux emerging from the volcano and reaching the observation point No. $4$ at $730$~m from the centre of geological structure.}
\begin{tabular}{lcccc}
\hline
\textbf{Points}        & \textbf{P$_{1}$}    & \textbf{P$_{2}$} & \textbf{P$_{3}$} & \textbf{P$_{4}$} \\ \hline
\textbf{Latitude  ($^{\circ}$N)}        & 4.49229             & 4.49198       & 4.48733       & 4.49494     \\
\textbf{Longitude ($^{\circ}$W)}        & -75.38109            & -75.38009      & -75.37951      & -75.38811    \\
\hline
\end{tabular}
\label{TableMachin}
\end{table}

As described in reference \citep{VesgaramirezEtal2020}, having considered technical and logistic data, we found that only Cerro Mach\'{\i}n can be feasibly studied through muography, and there we identified four observation points (shown in Table \ref{TableMachin}) which are not screened by any geological structures behind. Some of the manifestations of the volcanic activity of Cerro Mach\'in are the presence of fumaroles in the dome, permanent microseismicity, thermal waters flowing in the vicinity of the crater, geoforms of the volcanic building and a more significant presence of radon gas in the sector \citep{rueda2005erupciones}. 

\begin{figure}
\begin{center} 
{\includegraphics[scale=0.1]{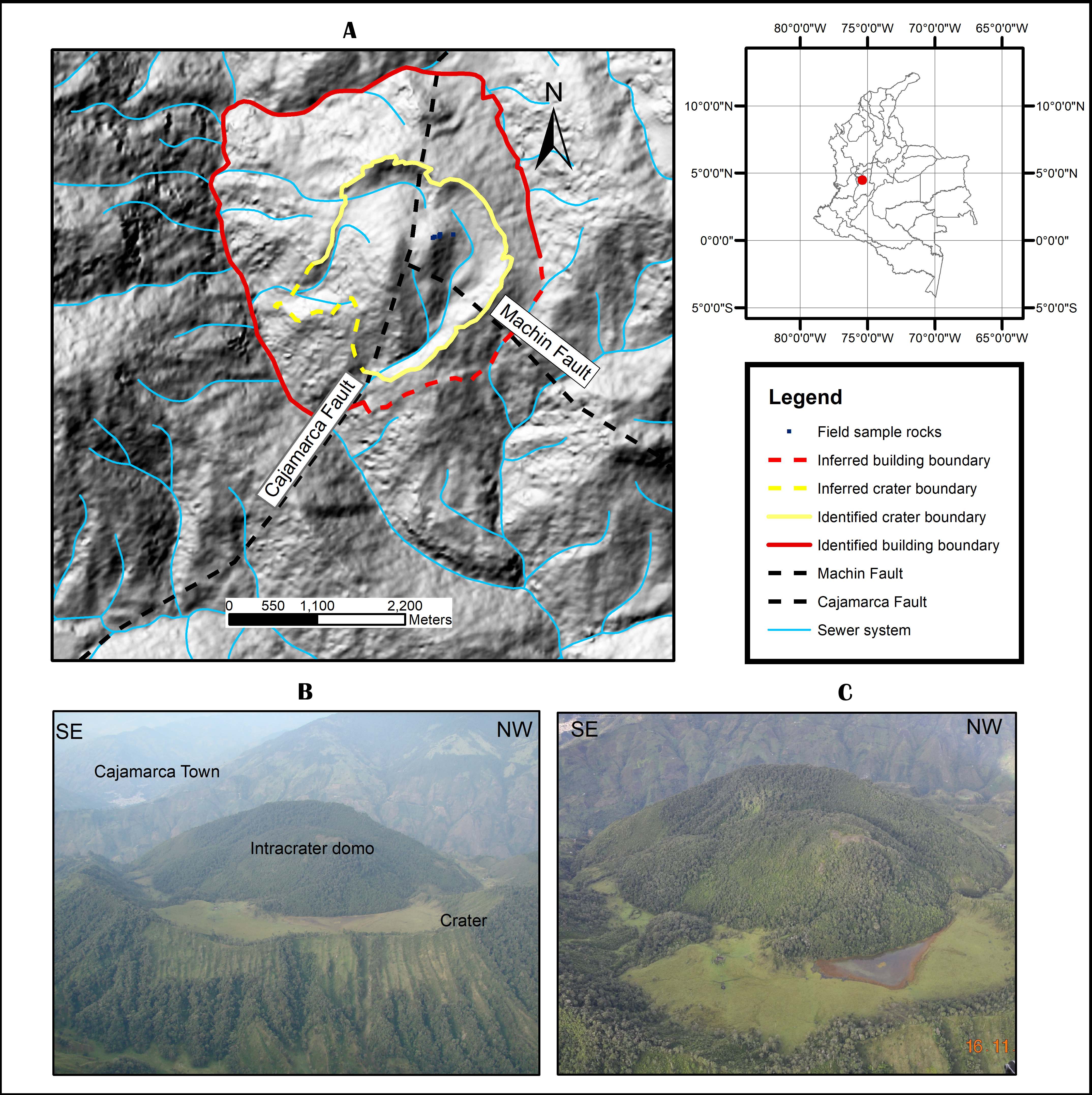}}
\caption{Location of the Mach\'in Volcano in the Central Cordillera of Colombia (red perimeter); the main route, the primary and secondary drains, the Cajamarca fault \citep{mosquera1978geologia} and the Mach\'in fault \citep{cepeda1995volcan} are also shown; B and C. VCM pictures. Note the intra-American dome in B and current intra-American “lagoon” in C (Pictures taken from the page of the Colombian Geological Service \url{http://www.sgc.gov.co}, 2011).}
  \label{LocationMachin}
\end{center}  
\end{figure} 
The Cerro Mach\'{\i}n volcano is in a strategic region on the eastern flank of  Colombia's Central Cordillera  (at 4$^{\circ}$29'23.08"N, $\;$ 75$^{\circ}$23'15.39"W), with a crater $2.4$ km diameter and $450$ m high dome (see figure \ref{LocationMachin}). It is one of the most dangerous active volcanoes in Colombia, having had six significant explosive eruptions in the last $5000$ years \citep{ laeger2013crystallization, cortes2001estudio, inguaggiato2017hydrothermal}. These eruptions have deposited many types of pyroclastic sediments with associated lahars that have travelled more than $100$~km  \citep{murcia20082500}. The last known Vulcanian eruption occurred $900$ years ago and produced associated pyroclastic flows.  This incident ended with the emplacement of an intra-crater dacitic dome and an active boiling fumarole field in the summit area of the central volcanic structure \citep{rueda2005erupciones, laeger2013crystallization, murcia20082500,  thouret1995quaternary}. Moreover, the recent increase of seismic activity recorded since $2000$ could result in a potentially higher threat to the neighbouring population \citep{sanchez2014riesgo}.

\section{The geological model and the forward modelling}
\label{GeologicalForwardModel}
To implement the geophysical inversion, we first devised a geological model for the density distribution and a plausible inner structure for the Mach\'in Volcano. Beginning with this improved representation of the inner volcano structure, we carried out the forward modelling to estimate the detected muon flux at a particular observation point.

\subsection {Cerro Mach\'in volcano density model}
\label{GeologicalModel}
The first step for implementing the forward modelling, discussed in the next section,  is to guess the inner density distribution for the geological structure. In this section, we present an estimation of the density distribution as well as the possible inner structure of the Mach\'in Volcano based on geological field information.

A geological model that accurately characterizes the spatial distribution of rock type, alteration, and structure of a geothermal system is the fundamental starting point for validating the inversion technique and testing ideas on the locations of potential fractures and permeable fluid pathways. We used GEOMODELER\footnote{\url{https://geomodelr.com/}} to design a model adapted to the Cerro Mach\'in topography. We also considered a geological survey carried out, 2D sections from interpreted geological maps \citep{cepeda1995volcan, mosquera1982mapa, piedrahita2018estratigrafia} and rock samples taken from the current crater, in the dome and the areas associated with fumaroles of the volcano. 

\begin{figure}
\begin{center} 
{\includegraphics[scale=0.08]{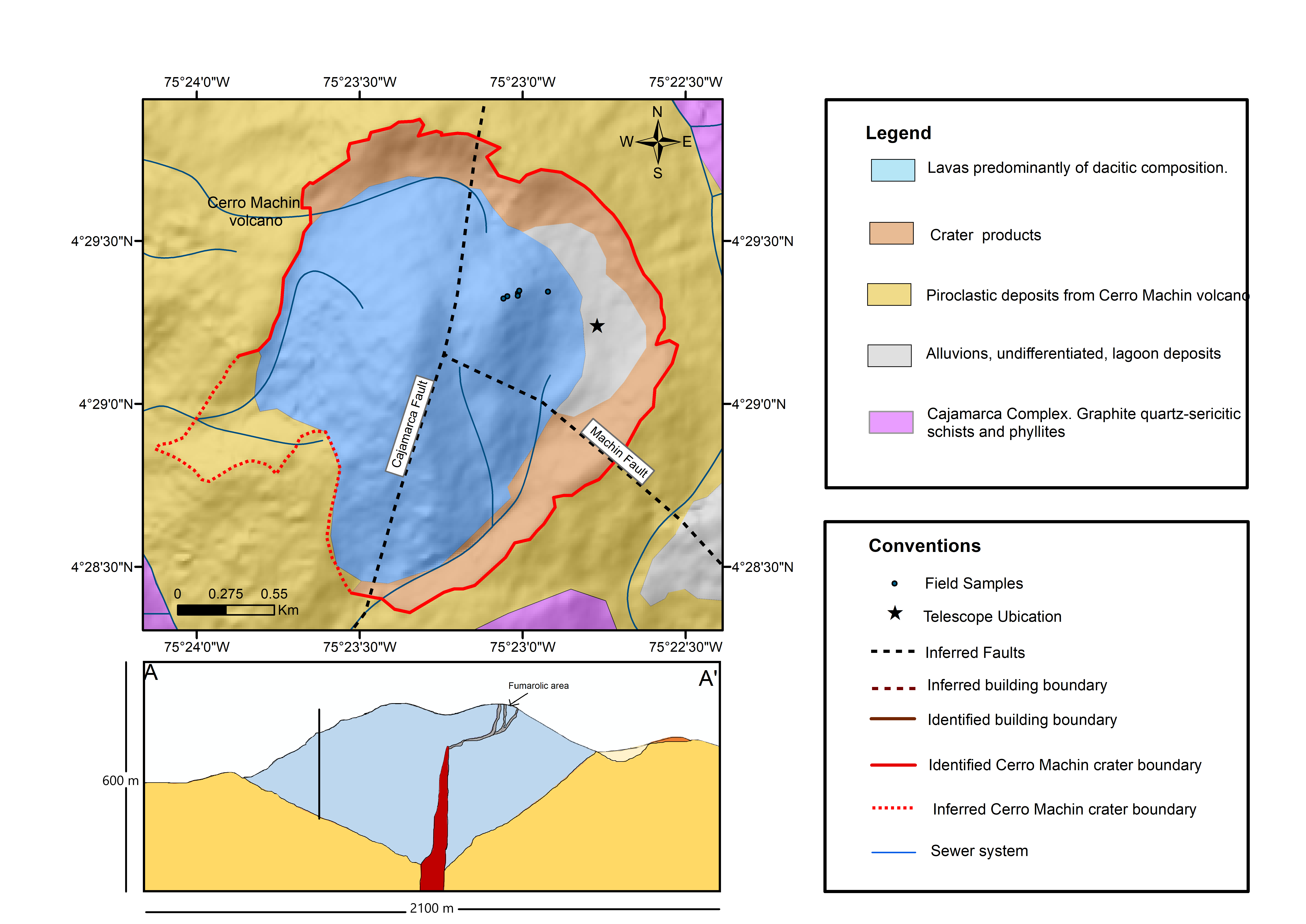}}\\
\caption[Geological map Cerro Mach\'in volcano] {Geological map of Cerro Mach\'in volcano and its dome.  In the lower part, we present a 2D profile in the A-A' direction, where the dimensions of the internal geological model of the volcano are displayed. 2D sections from interpreted geological maps \citep{cepeda1995volcan, mosquera1982mapa, piedrahita2018estratigrafia} and rock samples taken from the current crater, in the dome and the areas associated with fumaroles of the volcano.}
\label{geological}
\end{center}
\end{figure}

Figure \ref{geological} shows the geological model of Cerro Mach\'in  used to create the density model. The volcano is located at the intersection of the Cajamarca and Mach\'in faults (directed N20$^{\circ}$E and N42$^{\circ}$W, respectively) \citep{rueda2005erupciones} and built up on the metamorphic rocks of the Cajamarca Complex.

The Cajamarca complex contains orthogneisses, phyllites, quartzites, greenschists, graphitic schists and local marbles \citep{laeger2013crystallization, vargas2005new, villagomez2011geochronology}. The Cerro Mach\'in volcanic edifice consists of a ring of pyroclastic material with a diameter of $2.4$~km and includes a dacitic intra-crater lava dome \citep{murcia20082500}. 

We are mainly interested in the density contrasts that could exist in the dacitic dome of the Mach\'in because muons are not coming from below the horizon. To better understand the complex geology and structure of the region, other researchers have conducted integrated geophysical studies of Cerro Mach\'in. For example, in studies as described  \citep{inguaggiato2017hydrothermal, londono2011tomografia} it is estimated that the volcano has at least two magmatic chambers located at depths of $5$~km-$8$~km and $10$~km - $15$~km. However, since several faults cross the volcano, it is believed that there is an interaction between tectonic and volcanic activity \citep{londono2016evidence}. The chemical and isotopic compositions found in the thermal samples reported in  \citep{inguaggiato2017hydrothermal}  supports the existence of a heat source that induces vaporisation of shallow thermal fluids of meteoric origin, releasing vapour in the upper part of the dome and feeding boiling water to the springs at its base.

\begin{table}
\begin{center}
\caption{Densities of samples from Cerro Mach\'in dome. The prevalent type of rock in the Cerro Mach\'in Volcano dome is dacitic with a density of $2.50$~g/cm$^{3}$, decreasing in the $M2$ and $M3$ zone due to fumarole area. The reported densities of these samples were evaluated by using the modified methodology proposed by Houghton and Wilson \citep{houghton1989vesicularity}, based on Archimedes Principle for particles in the range of $32$~mm to $8$~mm in diameter.}

\begin{tabular}{|l|c|}
\hline
\multicolumn{1}{|c|}{Sample}                                                     & Density (g/cm$^{3}$) \\ \hline
$M1$ (dark grey, unweathered dacitic)                                                         & $2.50$             \\ \hline
$M2-A$ (dark grey, slightly weathered dacitic)                                                   & $1.83$             \\ \hline
$M2-B$ (light grey, slightly weathered dacitic)                                                 & $1.93$             \\ \hline
$M2-C$ (light grey, slightly weathered dacitic)                                                 & $2.11$             \\ \hline
$M3-A$~fragment (light beige, weathered dacitic)                                                 & $1.73$             \\ \hline
$M3-B$~fragment (light beige, weathered dacitic)                                                 & $2.18$             \\ \hline
\end{tabular}
\label{Tabledensity}
\end{center}
\end{table}

\begin{figure}[h!]
\begin{center} 
{\includegraphics[scale=0.5]{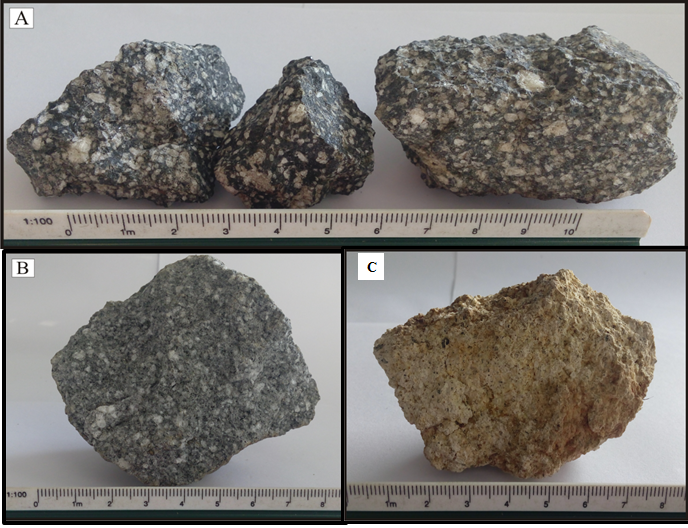}}
\caption[Rock Samples] {$M1$: A) Sample of a ``healthy rock'', its components such as plagioclase crystals and the vitreous matrix, are without evident alteration. B) $M2$: Sample with beginnings of mineral alteration and smaller crystals. C) $M3$: Altered rock sample; its components are no longer the original minerals (plagioclases), they were mainly replaced by clays.}
  \label{MachinRock}
\end{center}  
\end{figure}

Table \ref{Tabledensity} lists the density of rock samples taken from the hillside of the Mach\'in volcano and complementing our 3D geophysical density model. Accordingly, the lava dome has three main areas characterised by different densities, as shown in figure \ref{MachinRock}. An area $M1$ associated with unweathered dacitic rock with a density of  $2.50$~g/cm$^{3}$.  A second and third area ($M2$ and $M3$, respectively) with samples related to the fumarolic field with densities between $1.73$~g/cm$^{3}$ - $1.95$~g/cm$^{3}$. 

\begin{figure}[h!]
\begin{center} 
{\includegraphics[scale=0.4]{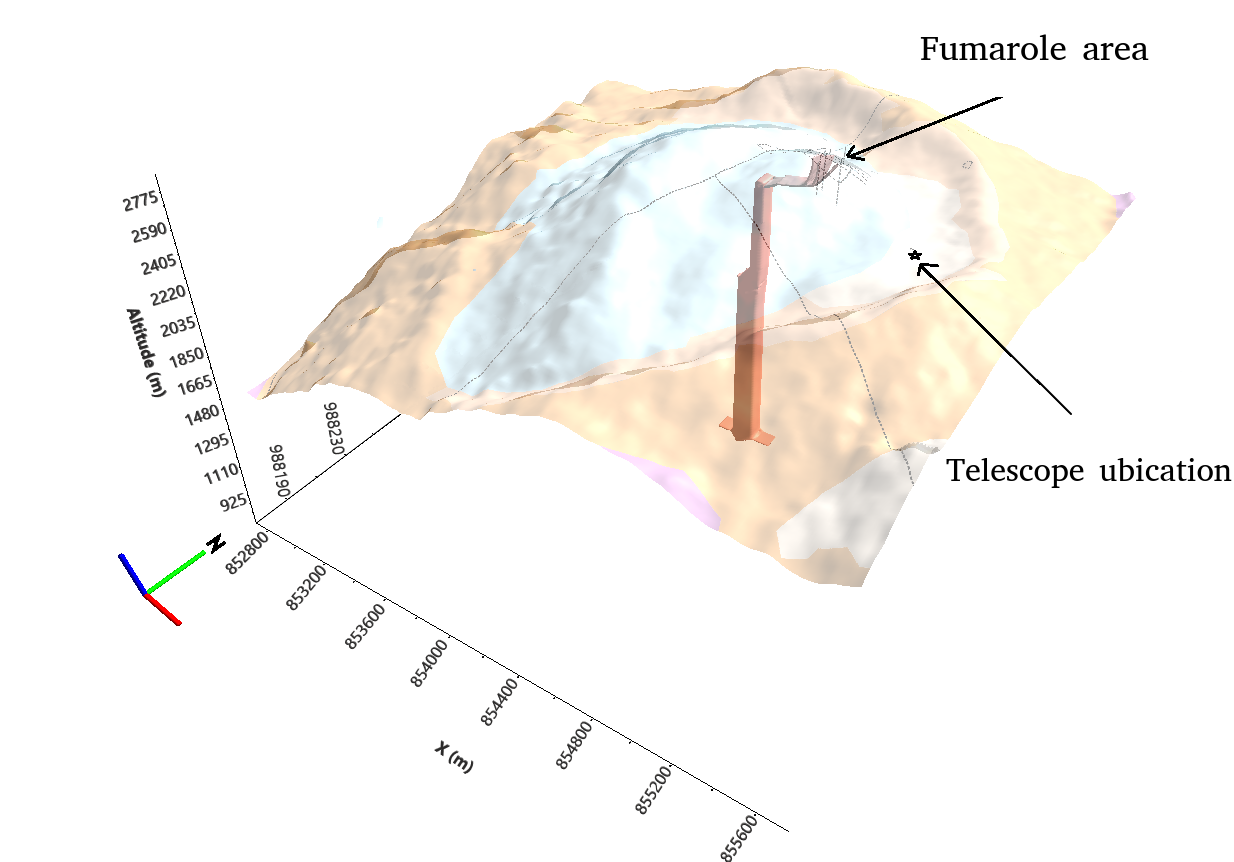}}
\caption[DensityModel] {The initial Cerro Mach\'in dome density model, to be used for our forward modelling test. The colours represent the internal volcano density distribution. The fumarole area with an average density of  $1.73$~g /cm$^{3}$ is where samples were collected, the blue zone denotes the predominant dome rock, dacite,  with a density of  $2.50$~g /cm$^{3}$, while the yellow zone represents the crater products rocks with a density of  $1.95$~g /cm$^{3}$. Finally, small black dots indicate the location of fumaroles and larger black squares shows the location of the muon telescope}
  \label{DensityModelMachin}
\end{center}  
\end{figure}

\subsection{Forward modelling}
\label{SectForwardModelling}
The forward modelling provides an initial estimation of the number of muons, crossing the volcanic structure (as described above in figure \ref{DensityModelMachin}), and impinging on the telescope. To determine the muon flux emerging from the Cerro Mach\'in volcano and impacting the instrument, we used three main elements: the open sky Reyna-Bugaev muon energy spectrum model \citep{lesparre2010geophysical, bugaev1998atmospheric, reyna2006simple}, the specific topography of the volcanic dome area, with the density distribution model and the energy losses schema for transmitted muons through matter.  The modelling consists of several steps described below. 

In the next sections we describe in details the forward modelling pseudo-code diplayed in the Appendix A, table \ref{ForwardModPseudo}.

\subsubsection{The open sky muon flux}
\label{OpenSkyFlux}
As an input we use the Reyna-Bugaev differential parametric muon flux model  \citep{lesparre2010geophysical, bugaev1998atmospheric, reyna2006simple},  

\begin{equation}
\Phi_{R}(\theta, p) = \cos{\theta}^{3} A_{R}\, (p \cos{\theta})^{-(\alpha_3 y^3 +\alpha_2 y^2 +\alpha_1 y +\alpha_0)} \, ,
\label{ReynaBagaevFlux}
\end{equation}
with $y = \log_{10}(p \cos{\theta})$ and  the momentum $p$ verifying,

\begin{equation}
pc^2 = E_0^2 -E_{\mu}^2 \quad \mathrm{where} \quad E_{\mu} \equiv m_{\mu} c^2 = 0.10566\; \mathrm{GeV}.
\label{EqnMomentum}
\end{equation}
The parameters considered are:  $A_R~=~0.00253$, $\alpha_0~=~0.2455$, $\alpha_1~=~1.2880$, $\alpha_2~=~-0.2555$ and $\alpha_3~=~0.0209$ \citep{reyna2006simple}. This semi-empirical and simple parametric model is valid for a wide energy range ($1$\;GeV $\leq E_0 \leq 2000$\,GeV) and zenith angles, and particularly for low energy muon fluxes.

In this paper, we use this simplified model for the atmospheric muon flux to focus on the density profile reconstruction using SAA. However, in other studies \citep{VesgaramirezEtal2020, AsoreyEtal2015B, AsoreyNunezSuarez2018, SarmientocanoEtal2020} we employed more accurate Monte Carlo simulations of muons using CORSIKA \citep{HeckEtal1998} and MagnetoCosmic \citep{Desorgher2003} codes to generate more realistic  atmospheric showers.

\subsubsection{Distances travelled by muons through the volcanic edifice}
The second stage is to estimate the distance travelled by muons in the volcano. This estimation includes the calculation of both the opacity of the material and the minimum energy required by muons to cross the geological structure.  The opacity of the material, determined by using the distances travelled by muons in the volcanic edifice, is associated with the mass density distribution $\rho$ integrated along the muon path $L$ as

\begin{equation}
\varrho \left( L \right) = \int_{L} \rho \left( \xi \right) d \xi=\bar{\rho} \times L,\
\label{opacity}
\end{equation}
where $\xi$ is a distance along the muon path through the geological edifice, $L$ is the total distance travelled by muons in rock and $\bar\rho$ the average density along the muon trajectory. 

We have modelled the muon energy loss along each path considering a uniform density distribution as

\begin{equation}
-\frac{dE}{d\varrho}=a(E)+b(E) E, \ 
\label{lostenergy} 
\end{equation} 
where $E$ is the muon energy, $a(E)$ and $b(E)$ depend on the rock composition/properties and $\varrho(L)$ is the density integrated along the trajectory of the muons (the opacity defined by equation (\ref{opacity})). As stated above, the prevalent type of rock in the Cerro Mach\'in Volcano dome is dacitic, mainly quartz ($\mathrm{Si} \mathrm{O}_2$) with a density of $2.50$~g/cm$^{3}$. 

The coefficients $a(E)$ and $b(E)$ are obtained from the report of the Particle Data Group \citep{OliveEtal2014}\footnote{Tables on: \url{ http://pdg.lbl.gov/2011/AtomicNuclearProperties/} }, where the function $a(E)$ represents the energy loss due to ionization, while $b(E)$ takes into account the contribution of radiative energy losses due to bremsstahlung, pair production and nuclear interactions. 

With the muon energy losses, it is possible to determine the minimum muon energy $ {E}_ {min}$ needed to cross a certain opacity $ X=\varrho$, given by

\begin{equation}
E_{min} = \int _{0}^{\varrho}  
\frac{ \mathrm{d} E }{ \mathrm{d}\varrho}  \mathrm{d} \varrho + E _{\mu}  \, , 
  \label{Emin}
\end{equation}
where $E _{\mu}$ is the muon rest energy, given in equation (\ref{EqnMomentum}). 

\begin{figure}[h!]
\begin{center} 
{\includegraphics[scale=0.51]{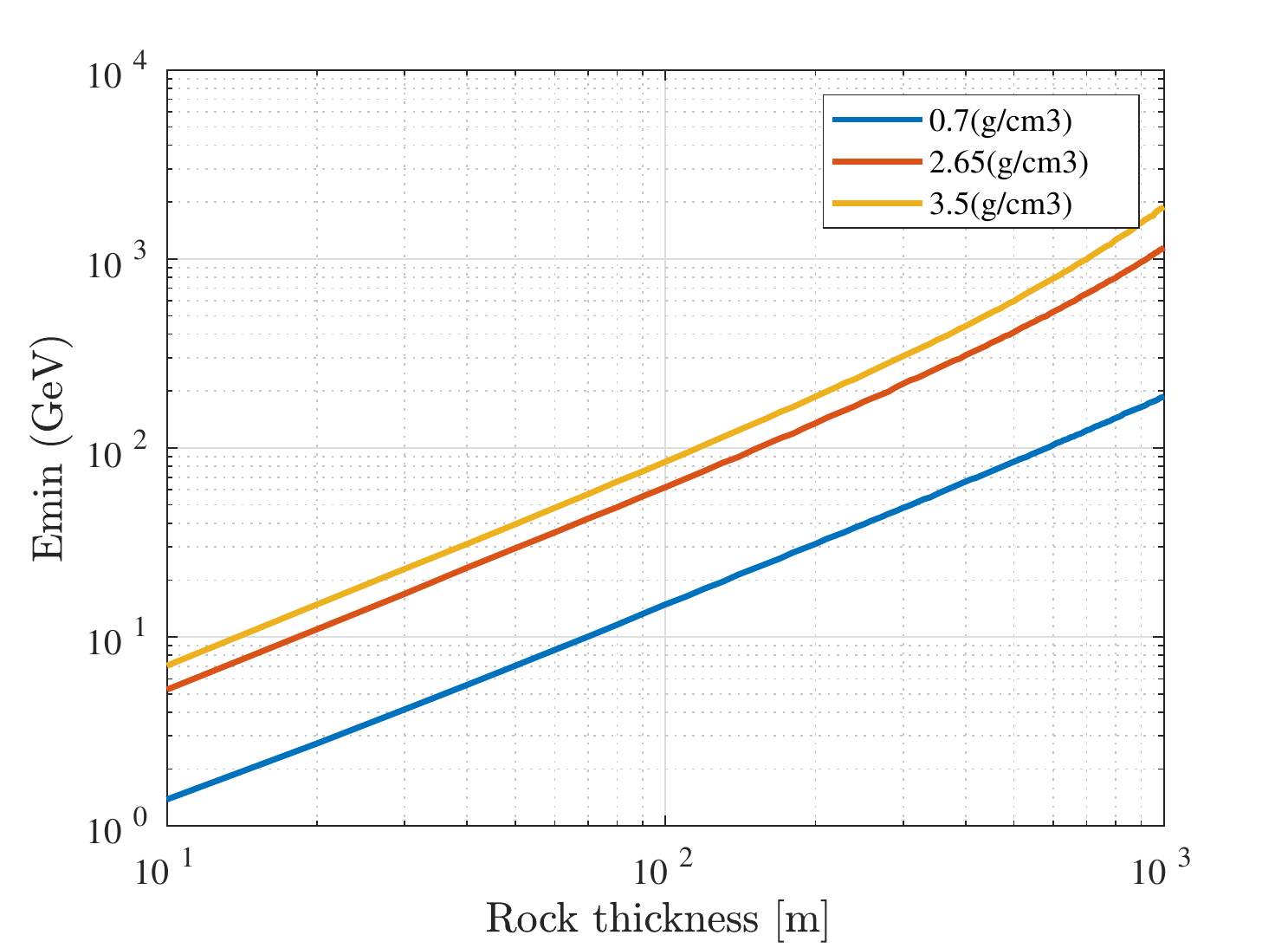}}
\caption[Minimum energy as a function of the rock thickness $L$] {The minimum energy of muons as a function of the rock thickness $L$. The three example densities are: $ 0.7$~g/cm$^{3}$, $2.65$~g/cm$^{3}$ and $3.5$~g/cm$^{3}$ respectively. A muon with an initial energy of $11.6$~GeV can cross a distance of $21$~m in standard rock, and for the energy of approximately $ 1.14~\times~10^{3}$~GeV, the maximum distance travelled by the muons  correspond to 996 meters. Similarly, it can be observed that for a density of $3.5$~g/cm$^{3}$, it takes almost three times as much energy to cross the same distance as that for a  $0.7$~g/cm$^{3}$.}
  \label{Emin_plot}
\end{center}
\end{figure}

Finally, solving this equation we obtain the minimum energy necessary for a muon to cross a given thickness of rock (see figure \ref{Emin_plot}). This model largely agrees with detailed simulations of muon flux in atmospheric showers and accurate muon transport through the rock  \citep{lesparre2010geophysical}. 

\subsubsection{The integrated muon flux}
The integrated muon flux, $I$, after crossing a certain thickness of rock, can be estimated from the differential muon flux $\Phi_{R}$ as

\begin{equation}
I=\quad \int _{ E_{min}(\varrho) }^{ \infty  }{ \Phi_{R}~\mathrm{d}{ E } } \quad ( \mathrm{cm}^{-2}~\mathrm{sr}^{-1}~\mathrm{s}^{-1}).
\label{flujointegrado}
\end{equation}

\begin{figure}[h!]
\begin{center} 
{\includegraphics[scale=0.5]{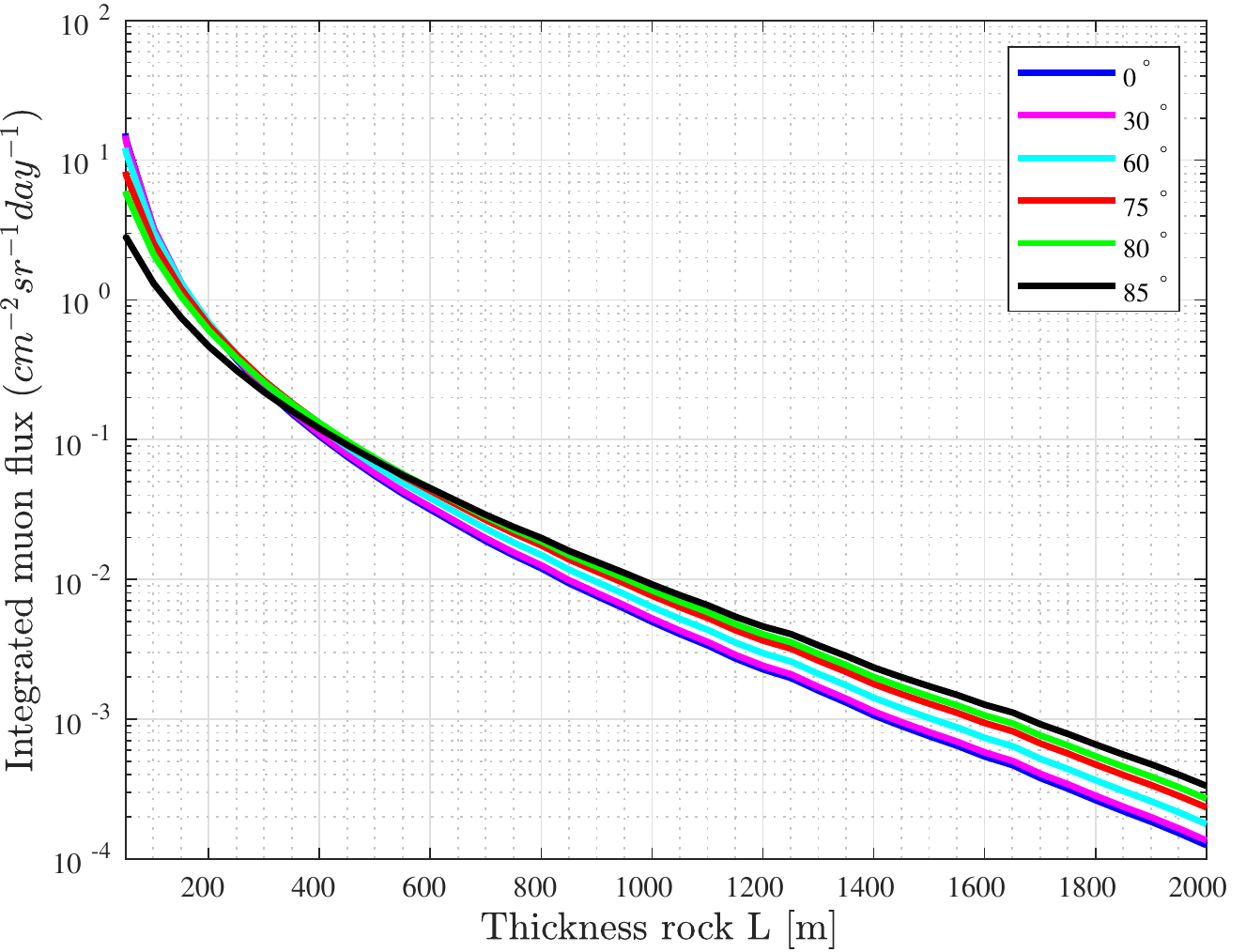}}
\caption[Integrated flux of muons for different zenith angles depending on the thickness of the rock.] {Integrated muon flux for different zenith angles depending on the rock thickness. Considering a rock thickness less than $800$~m, $30^{\circ}$ of zenith angle and $100$~days of recording time, we can obtain $\sim 0.4$~muon/cm$^2$, and consequently, $\sim 5000$  muons detected, during this period.}
  \label{Integrated_flux}
\end{center}  
\end{figure}

Figure \ref{Integrated_flux} displays the integrated muon flux $I$ impinging on the detector after passing through a pre-defined thickness of rock in a volcano at different zenith angles. We have chosen to consider the rock thickness less than $1000$~m to ensure a statistically significant measurement of muon flux over a period of a few months.

\subsubsection{The instrument acceptance}
\label{AceptanceNumbers}
Finally, detailed knowledge of the instrument capabilities is essential in estimating the number of muons which cross the structure and impact the telescope \citep{lesparre2012density}. The acceptance function ($\mathcal{T}$, measured in cm$^2\cdot$ sr) is one of the key features to take into account because it converts the integrated flux $I$ onto the number of detected  muons $N$ as  

\begin{equation}
N(\varrho)= \Delta t \times \mathcal{T} \times I(\varrho), 
\label{Nmuons}
\end{equation}
where $I\equiv I(\varrho)$ is the integrated flux (measured in cm$^{-2}\cdot$sr$^{-1}\cdot$s$^{-1}$), $\mathcal{T}$ the acceptance function (measured in cm$^{2}\cdot$sr), $\Delta t$ the recording time, and $\varrho~=~\varrho(L)$ represents the opacity parameter. 

Let us consider two impacted pixels in the two panels by the same impinging muon. In the front panel the pixel is labeled as $P^{F}_{(i,j)} = r^{F}_{(i,j)}$ and in the rear panel as $P^{R}_{(k,l)} = r^{R}_{(k,l)}$. The subscripts $(i,j)$ and $(k,l)$ indicate the position in each detector matrix of $N_x~\times~N_y$~pixels and range from $r^{F/R}_{0,0}$ found at one corner to $r^{F/R}_{29,29}$ located at the opposite corner. Thus, we can identify $(2N_x~-~1)(2N_y~-~1)$ different particle trajectories, $r_{m,n}$, shared by the same relative position, $m~=~i-k$ and $n~=~j-l$  \citep{lesparre2010geophysical}.

The acceptance is obtained multiplying the detection area, $S(r_{m,n})$, by the angular resolution, $\delta\Omega(r_{m,n})$, i.e.

\begin{equation}
\mathcal{T}(r_{m,n})=S(r_{m,n})\times \delta\Omega(r_{m,n}),
\end{equation}
where $r_{m,n}$ represents each discrete muon incoming direction.  

The acceptance and the corresponding number of impinging muons detected are a function of the telescope's geometrical parameters such as the number of pixels in the panels ($N_x~\times~N_y$), size of pixels ($d$), and separation of detection surfaces ($D$).

\begin{figure}[h!]
\begin{center} 
{\includegraphics[width=0.45\textwidth]{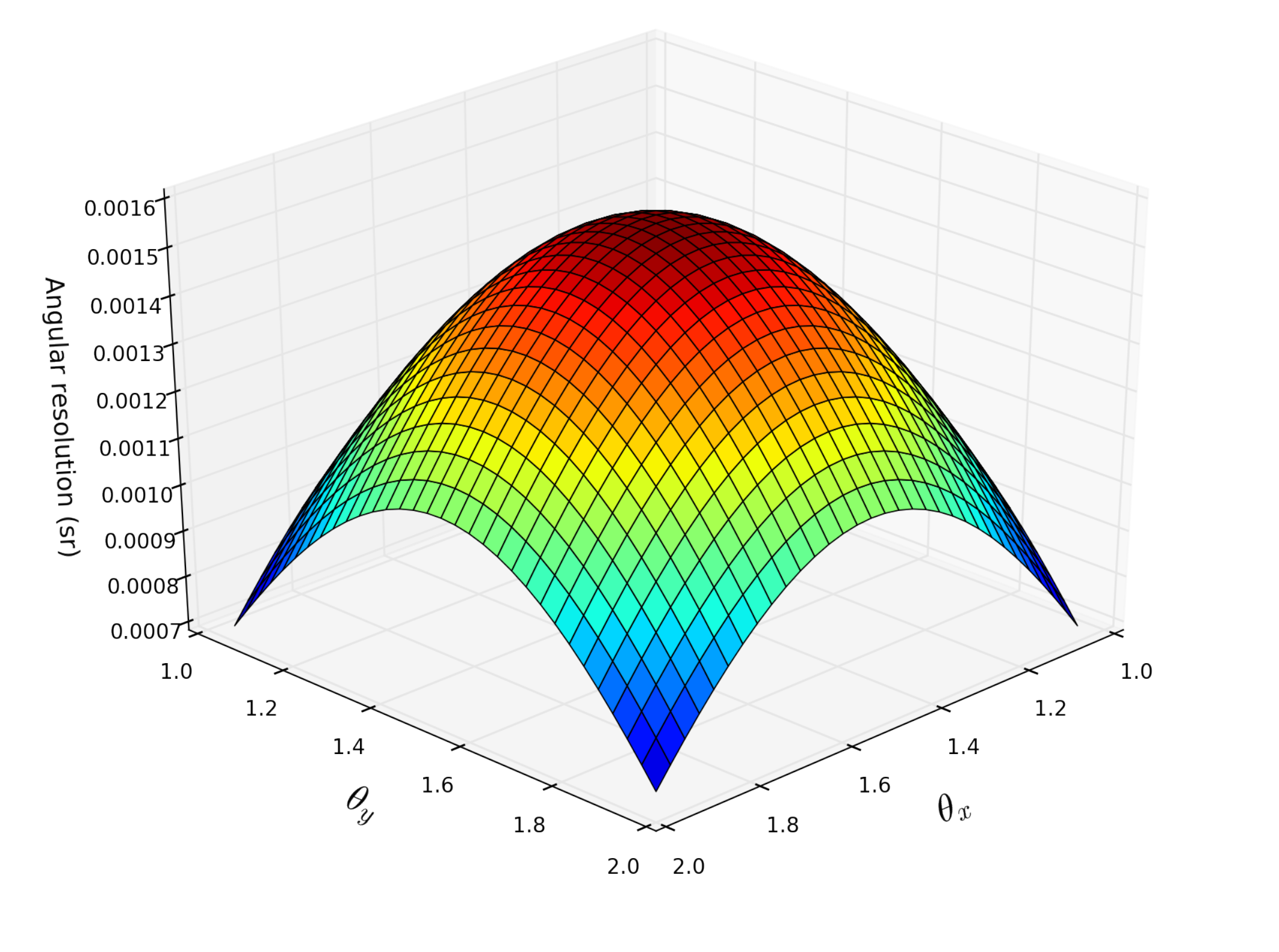}}
{\includegraphics[width=0.45\textwidth]{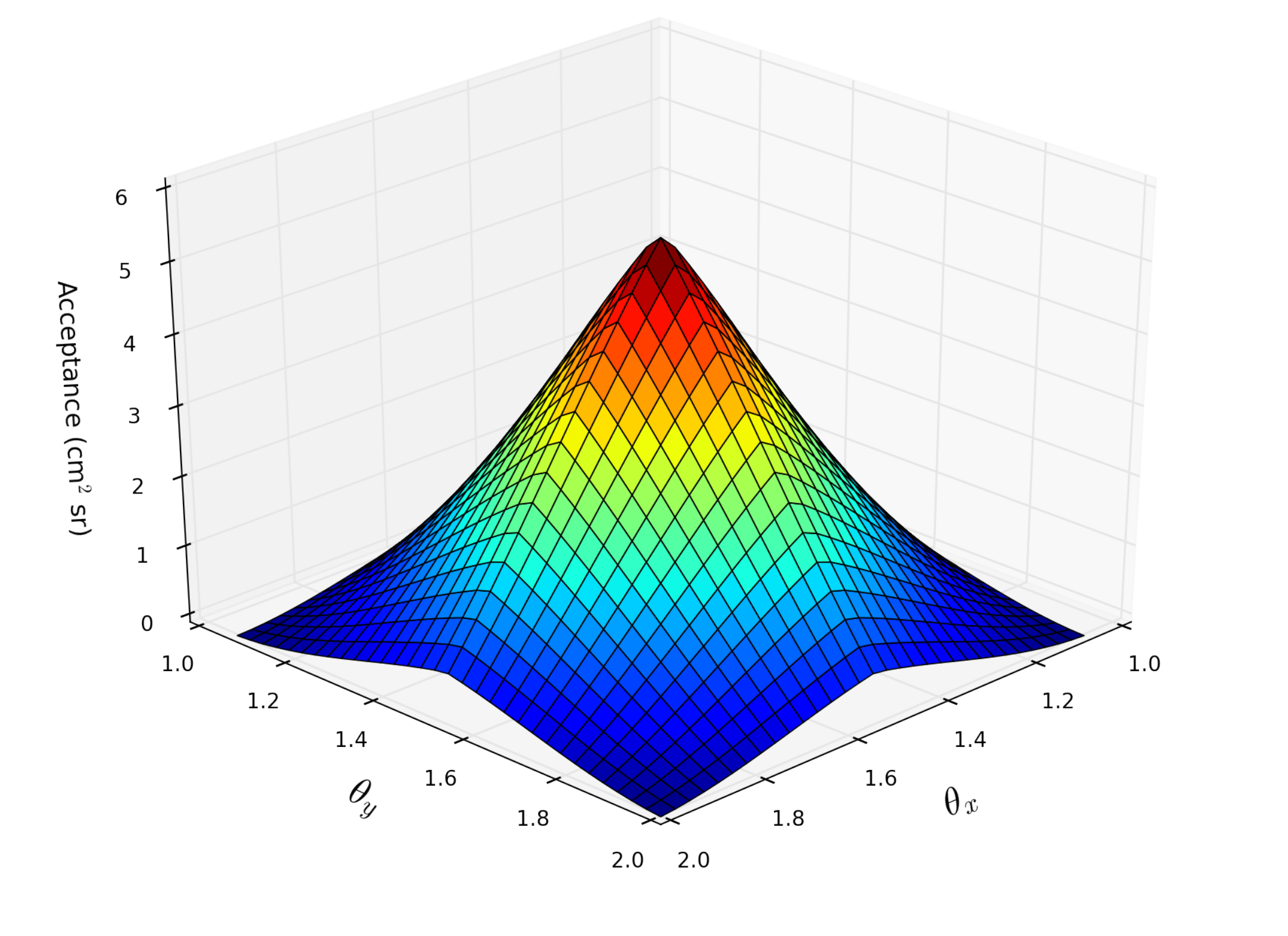}}
\caption[Angular resolution and acceptance function] {Angular resolution (sr) and acceptance function ({cm}$^{2}$~sr) for the MuTe project, with a separation of $D =200$~cm  between both panels. Each detection panel has $N_x = N_y = 30.$ $4$~cm  wide scintillation bars, providing $900$ pixels of $16$~cm$^{2}$ detection area. There are $3481$ possible discrete incoming addresses $r_ {mn}$, for a maximum acceptance function of $5.759$~cm$^{2}$~sr for the MuTe project.}
  \label{acceptance1}
\end{center}  
\end{figure}

Figure \ref{acceptance1} displays the angular resolution and acceptance function for the MuTe hodoscope with $N_x~=~N_y~=~30$ scintillator bars, size of pixel $d~=~4$~cm  and $D~=~200$~cm.  The total angular aperture of the telescope (considering both panels and the WCD) with this configuration is roughly $50^{\circ}$ ($0.9$~rad) with a maximum solid angle of $1.024~\times~10^{-3}$~sr at $r_{0,0}$, and the largest detection area approximating $5.759$~cm$^{2}$~sr at $r_{0,0}$ (see reference \citep{PenarodriguezEtal2020} for details).

\subsubsection{The numbers of detected muons}
As shown in the next section, the input for  the geophysical inversion should be the distribution of observed muons. Therefore, for the present work we use as ``observed'' number of muons those coming from a simulation described in the forward modelling having a known inner density configuration. 

\begin{figure}[h!]
\begin{center} 
{\includegraphics[width=0.7\textwidth]{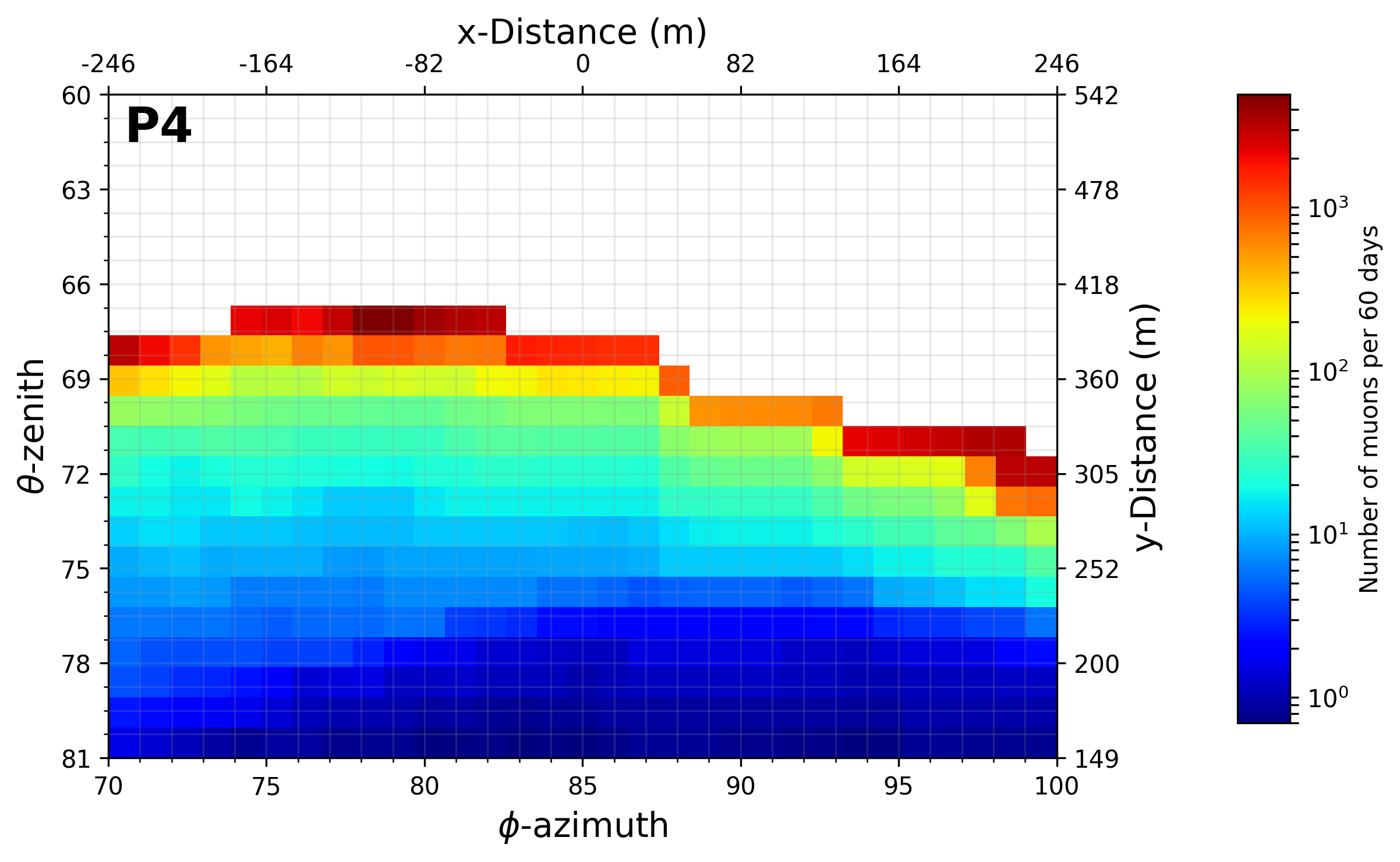}}
\caption[Detected muon number] {Left panel of the figures displays the muon flux crossing the volcano and collected at the observation point No. $4$ during  $60$ days.}
  \label{NumbersMuons}
\end{center}  
\end{figure}

In figure \ref{NumbersMuons} we present the distribution of $N_{obs}$, collected at the observation point No. $4$, from all possible directions, during a time-lapse of $\Delta t \approx 60$ days. The average muon flux obtained is $\sim~150$~muon/pixel, reaching $\sim~100$~muon/pixel from the centre of the volcanic structure. If we define the maximum observed depth where the emerging muon flux is less than 10$^{-2}$ muons per cm$^{2}$ per day --with a zenith angle $\theta \approx 82^\circ$-- then we can only explore inner structures up to the depth of $\sim 190$~m from the top of the volcano \citep{VesgaramirezEtal2020}. 

\section{The geophysical inversion}
\label{InverseModelling}
Different studies \citep{lesparre2012bayesian, davis2011joint, rosas2017three, lelievre2019joint} discuss the problem of geophysical inversion in muography in combination with other complementary techniques. In this work we present an inversion implementation of muography using the SAA, to find an improved inner density distribution of Cerro Mach\'in volcano. 

The central point with any inversion method is to enhance the modelling with measured information, expressing the differences between the recorded and modelled data in terms of a real number known as a \textit{cost}.  The idea of this strategy is to find the best model (i.e. a \textit{solution}), improved by the measured data, which minimizes the cost function.

There are several ``local'' optimization strategies, schemes like least squares and gradient methods, which start from an initial model and look for other neighbouring solutions.  The main problem with the local methods is the possibility to get stuck in a bounded minimum where the cost function does not vary significantly. 
In contrast, global optimization algorithms systematically explore the variety of possible neighbouring solutions, decreasing the probability in identifying a local minimum instead of a global one. Monte Carlo simulation, genetic algorithm, particle swarm optimization and simulated annealing, search for a minimum value of the cost function choosing new solutions in a stochastic way.

In this work, we adopted SAA because of its simplicity, and excellent results in a wide range of disciplines \citep{kirkpatrick1983optimization, tarantola1982inverse,  gibert1991electromagnetic, pessel2003multiscale}.  For problems when the solution space is small, simulated annealing has a faster time with respect to another global optimization methods \citep{otubamowo2012comparative}.  The name of simulated annealing arises from the physics of annealing in solids when a crystalline solid is heated and later allowed to a prolonged cooling. The final configuration of the solid has the most regular possible crystal lattice layout with minimal energy.

In table \ref{InverseModPseudo} within Appendix B, we present the pseudo-code to implement the geophysical inversion applying SAA.  We have replicated the idea of solid annealing with the \textit{physical temperature} $T$, translated into the \textit{evolution parameter} and implementing the \textit{cooling process} by decreasing $T$ in predefined $\Delta T_{k}$ steps.  

The corresponding \textit{state of the physical system}, is defined by the average density distribution function for all zenithal \& azimuthal angles in the observation range. The associated muon number,  $N_{sim}(\bar{\rho})$, and the corresponding cost function $E(\bar{\rho})$,  implies a candidate \textit{solution} to the problem.  

As stated above, the input data to SAA is the number of observed muons, $N_{obs}$, collected at the observation point No. $4$ during  $60$ days and displayed in figure \ref{NumbersMuons}. Then the method starts by choosing the initial evolution parameter, $T_{initial}$, the \textit{cooling} scheme, $\Delta T_{k}$, and the initial random average density, $\bar{\rho}$. 

It is critical to identify the appropriate initial evolution parameter $T_{initial}$ and to select it. We have followed the statistical approach discussed in reference  \citep{weber2000seismic}, which can be written as:

\begin{equation}
T_{initial} \geq \sqrt{\overline{E(\bar{\rho}_{r},T)^2} - \overline{E(\bar{\rho}_{r},T)}^2} \, ,
\label{Tinitial}
\end{equation}
where $\overline{E(\bar{\rho}_{r},T)^2}$ stands for an average of the square of the cost function for different $T$s and $\bar{\rho}_{r}$ models, while $\overline{E(\bar{\rho}_{r},T)}^2$ represents the square of the average of cost function for the same sample of $T$s and $\bar{\rho}_{r}$s. In Appendix C, we sketch a modified SAA  to calculate these averages from different values of $T$s and $\bar{\rho}_{r}$s. The algoritm generating samples to select the initial parameter, $T_{initial}$, is essentially the SSA, but suppressing all executions loops.

The \textit{cooling} scheme, $\Delta T_{k}$, i.e. the way the \textit{temperature} decreases, is also crucial and we set it as 

\begin{equation}
\Delta T_{k} = -0.01\, T_{k} \quad \Leftrightarrow \quad 
T_{k+1} = 0.99\, T_{k}\, ,
\label{DeltaT}    
\end{equation}
which is very common in several other SAA implementations \citep{peprah2017optimal, mahdi2017performance,  cabrera2014dynamically, khairuddin2019comparison}. 

The initial random average density, $\bar{\rho}$ --for all possible emerging volcano trajectories-- is taken from a list of $201$ equal spaced values within the typical density range for volcanic rocks, i.e. $1.5$~g/cm$^3$ to $3.5$~g/cm$^3$. Next, we estimate the simulated number of muons $N_{sim}(\bar{\rho})$ impacting each pixel of the telescope  and calculate the initial cost function $E~=~\left\| N_{ obs }- N_{sim}(\bar{\rho}) \right\| $, the difference between the number of muons ``observed'' $N_{obs}$, and those obtained from the initial simulated data $N_{sim}(\bar{\rho})$; where $\left\| . \right\|$ denotes the  $L^{2}$ norm of a vector space.  

Now, for each step $\Delta T_{k}$ in the evolution parameter $T$, we generate a random neighbour state of the system for all azimuth and zenith angles in the observation range. Then we calculate an random average density, $\bar{\rho}_{r}$, its associated number of muons $N_{sim}(\bar{\rho}_{r})$ and the corresponding cost function for this neighbouring model,  $E(\bar{\rho}_{r},T)~=~\left\| N_{obs}~-~N_{sim}(\bar{\rho}_{r}, T) \right\|$.  

If $\Delta E(\bar{\rho}_{r}, T)~=~E(\bar{\rho}, T)~-~E(\bar{\rho}_{r},T)$, the difference in the cost function for this neighbour model is less or equal to zero, we accept the system state, i.e. the $\bar{\rho}_{r}$, its resulting muon number, $N_{sim}(\bar{\rho}_{r}, T)$, and the associated cost function $E(\bar{\rho}_{r},T)$ as a candidate solution; otherwise the model is not rejected but admitted with a probability given by the Metropolis's statistical criterion \citep{metropolis1953equation}. Thus, we calculate the probability  

\begin{equation}
P= \mathrm{exp}\left( -\frac{\Delta E(\bar{\rho}_{r}, T)}{ T} \right)
\label{MetropolisProbability}
\end{equation}
compare it with a random number $0 < r_{discrim} < 1$ and accept the new state of a system if $P>r_{discrim}$ and if not,  reject it. 

Before decreasing the evolution parameter, $T$, we perform several iterations to refine the obtained model based on the Metropolis' criterion. Next, $T$ is reduced, and the cycle is repeated until reaching the final value, typically: $T_{final} \rightarrow 0$.

\section{Implementing the inversion}
\label{ImplementingInversion}
\begin{figure}[h!]
\begin{center}
{\includegraphics[width=0.49\textwidth]{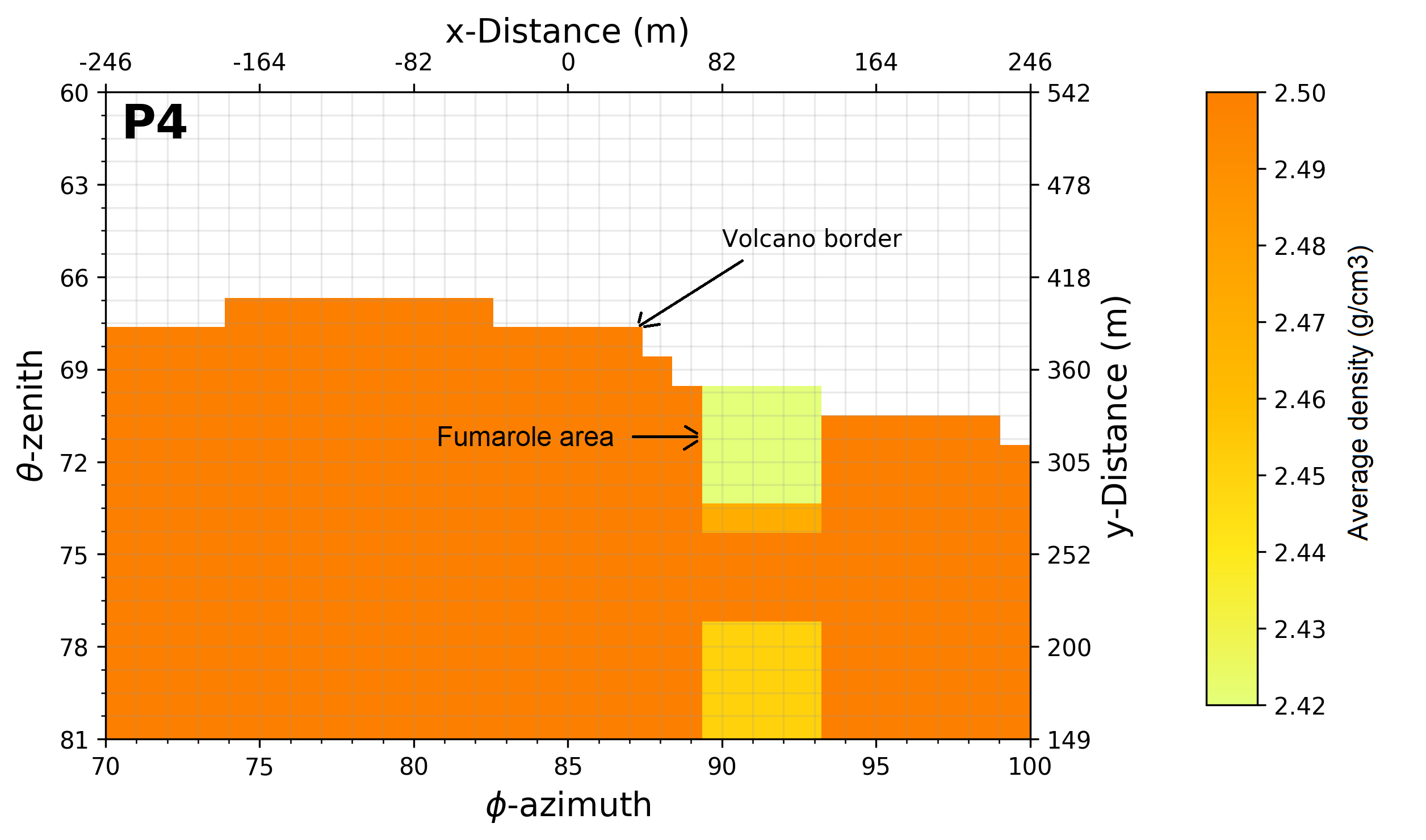}}
{\includegraphics[width=0.49\textwidth]{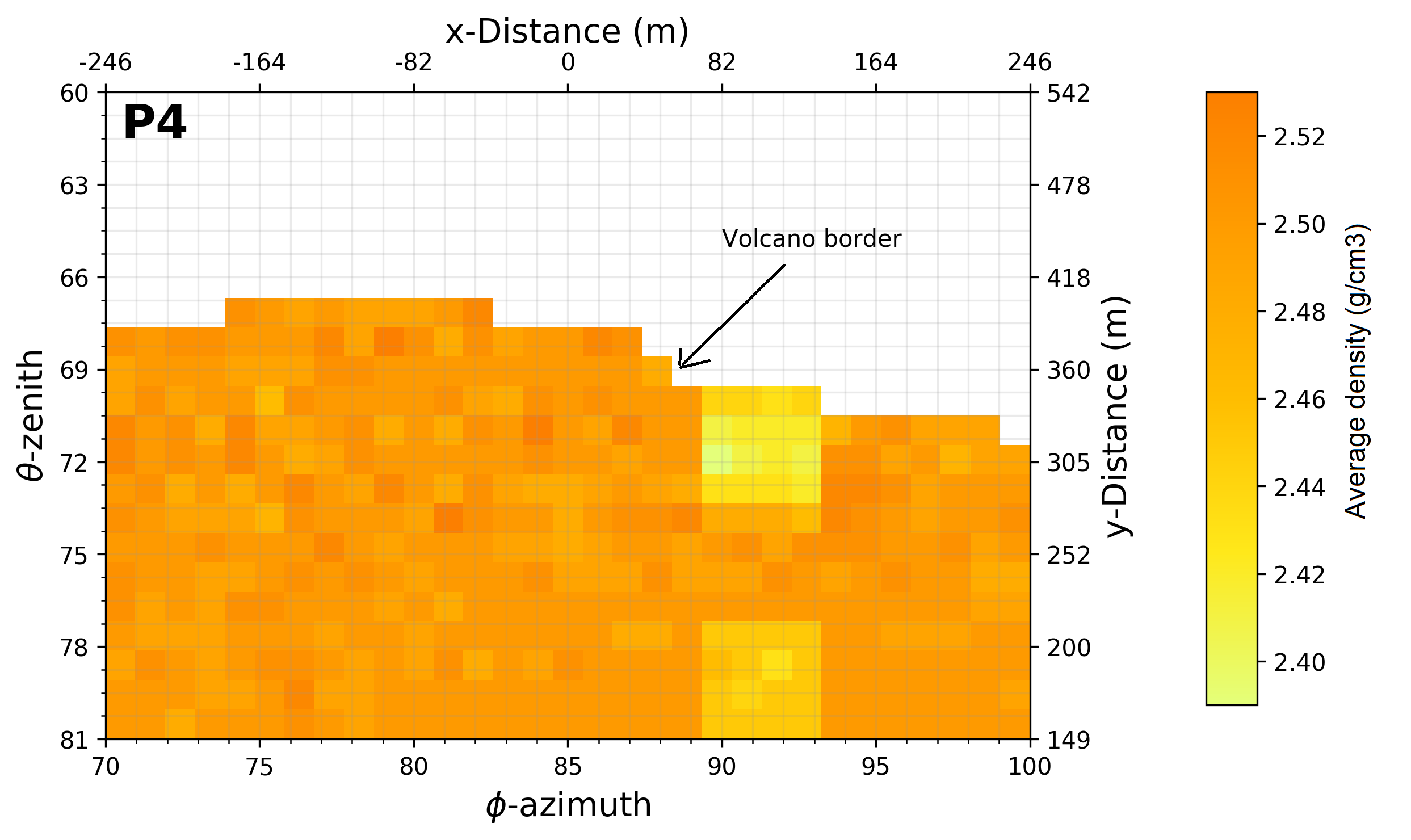}}
\caption[Before and after inversion]{On the left, the synthetic density model, and on the right, the profile after the inversion. The contrast of densities between volcanic conduit and the surrounding rock is maintained.}
  \label{DensityInversionMachin}
\end{center}  
\end{figure}

As discussed in section \ref{GeologicalForwardModel}, shown in figure \ref{DensityModelMachin}, and in the left plate of figure \ref{DensityInversionMachin}, the proposed Cerro Mach\'in density model consists of a dacitic lava dome with an active boiling fumarole chimney in the summit area of the central dome with its fumarole duct. As for the density model, we have divided the dacitic dome into cubes ($12.5$~m $~\times~12.5$~m $~\times~12.5$~m) using a digital elevation model with a resolution of $12.5$~m . The duct of $40$~m$^{2}$, with a density of $0.50$~g/cm$^{3}$, could contribute to the gas escape in the $M2$ zone.

\begin{figure}[h!]
\begin{center} 
{\includegraphics[width=0.4\textwidth]{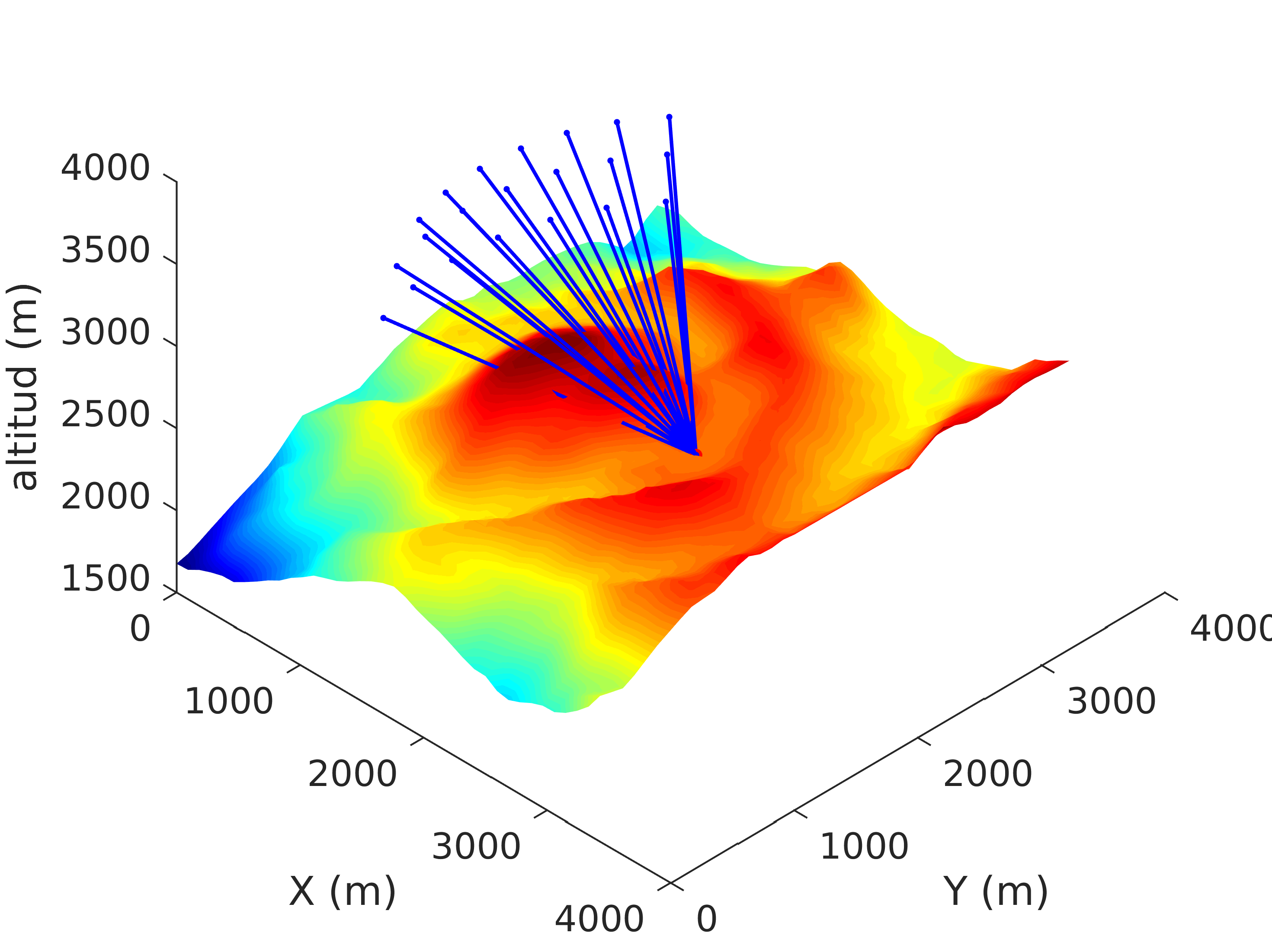}}
{\includegraphics[width=0.45\textwidth]{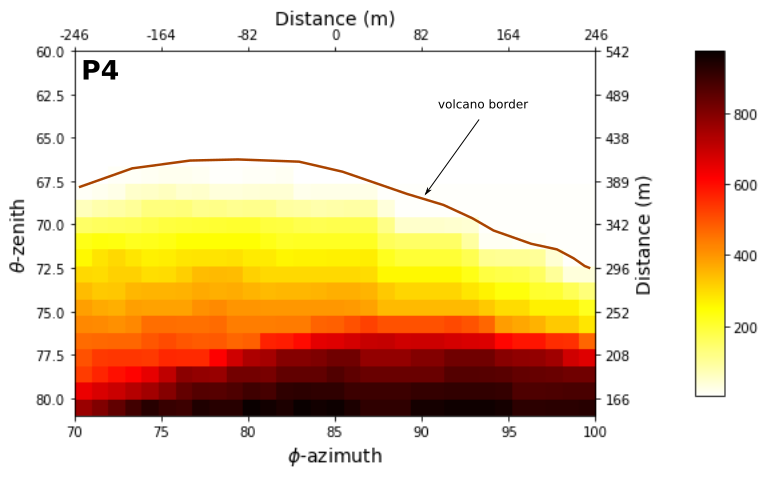}}
\caption[Trajectories of particles crossing the dome] {Left panel of the figures displays the muon flux crossing the volcano and collected at the observation point No. $4$ during  $60$ days. In the right panel we represente the distances travelled by particles through the dome of volcano (right) measured at the same observation point. Note that for this  point the distances in rock do not exceed $1200$~meters. The topographic data to calculate both plots come from the Global Digital Elevation Model of the Earth generated by NASA through the Mission Shuttle Radar Topography $380$, with resolution SRTM3 $1$.}
  \label{DistancesP4}
\end{center}  
\end{figure}

With the above geophysical density distribution pattern, our forward modelling was implemented simulating the muon flux data crossing and reaching  observation point No. $4$ at the base of the Cerro Mach\'in, $730$~m from the centre of the volcanic edifice (see figure \ref{DistancesP4}, Table \ref{TableMachin} and reference \citep{VesgaramirezEtal2020}). The thickness of rock traversed by muons as a function of the direction of arrival is shown in figure \ref{DistancesP4}, where it is seen that the maximum length travelled by muons is $L_{max}\approx 1200$~m.

The ``typical observation'' time scale is  $\Delta t \approx 60$ days, with the $30~\times~30$ pixel hodoscope, having an inter-panel distance of $200$~cm. We set a minimum statistic threshold count of $100$ muons/pixel so as to obtain images with a resolution of $59~\times~59$ pixel.  Notice that there are regions where the incoming muon flux is absorbed due to the volcano's geometry, and some of the distances travelled could easily exceed  $900$ meters. The statistics of $100$ muons/pixel for the above MuTe configuration is consistent with other more realistic studies  ( see references \citep{VesgaramirezEtal2020,  VasquezEtal2020, PenarodriguezEtal2020, MossEtal2018} for further details). 

From equation (\ref{DeltaT}), we set $T_{k+1} = 0.99\, T_{k}$ and from (\ref{Tinitial}) we choose $T_{initial} = 0.004$ obtained from a sample of $15$ cases obtained after executing the modified SAA sketched in Appendix C.

\begin{figure}[h!]
\begin{center} 
{\includegraphics[scale=0.5]{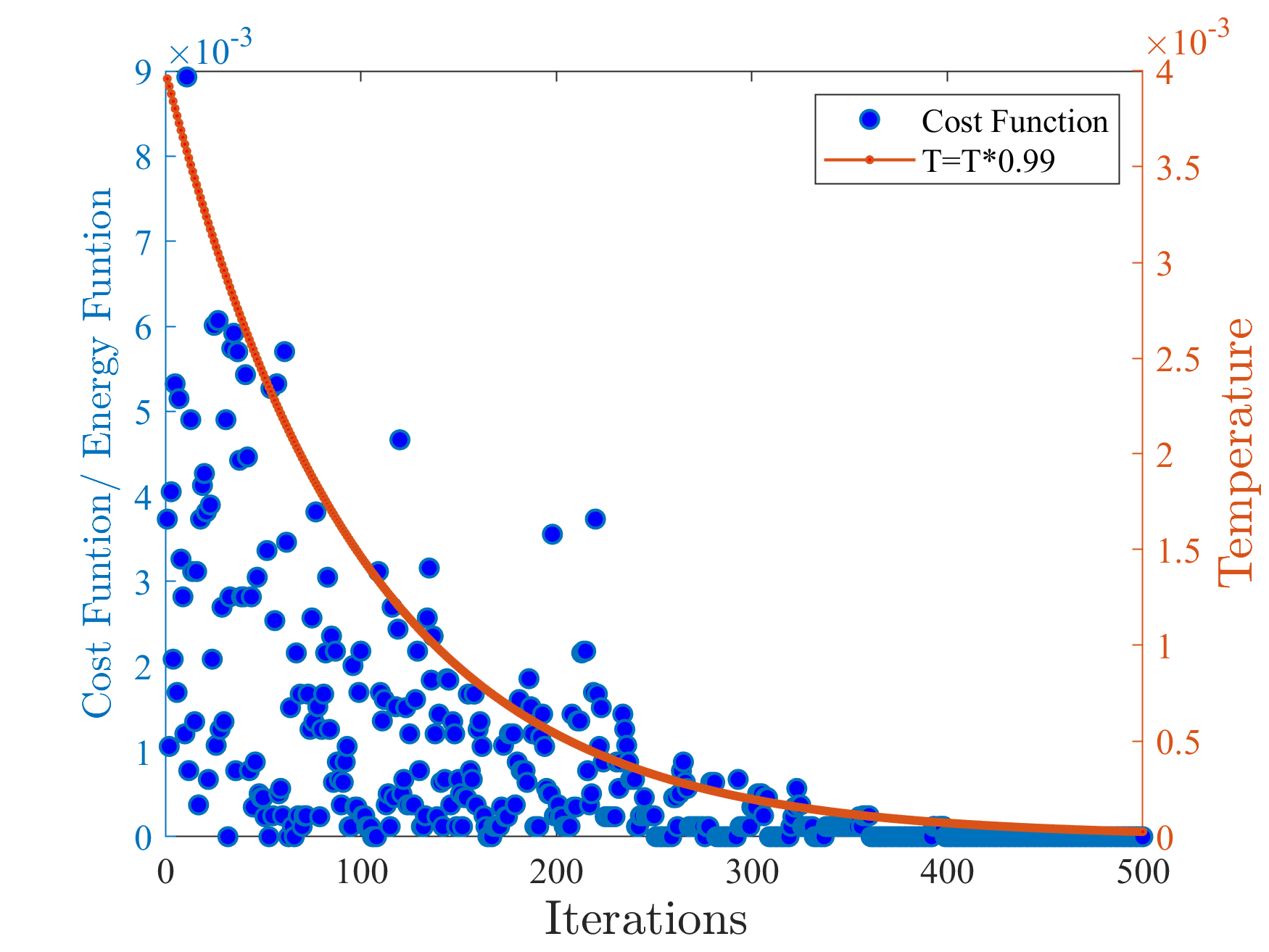}}\\
\caption[temperature] {Minimization of the cost function with SAA for one-pixel muography. The SAA loop starts at $T=0.004$, follows a ``cooling'' route implemented by  $T_{k+1}=0.99 \times T_{k}$, and obtains the best solution of density profile for the muon observed flux as $T \rightarrow 0$. Each dot has an additional refinement of $10~\times$ in the SAA execution scheme. }
\label{Temperature}
\end{center}  
\end{figure}

Then, we performed the geophysical inversion following the SAA obtaining the best density profile, which minimizes the cost function between the simulated and the ``observed'' muon numbers. In figure \ref{Temperature} we display the evolution of the optimization process, i.e. the decrease of the cost value --as well as the ``temperature'' parameter $T$-- as a function of the number of iterations. It is worth mentioning that each blue dot has an additional refinement of $10\times$ in the SAA execution.

The internal density models of Cerro Mach\'in volcano before and after the inversion are shown in figure \ref{DensityInversionMachin}. The left plate illustrates the initial density profile as seen from  observation point No.4. It is a density model consisting of a dacitic lava dome with a density of $2.50$~g/cm$^{3}$, and an active boiling fumarole duct of $40$~m$^{2}$, with a density of $0.50$~g/cm$^{3}$, (see section \ref{SectForwardModelling}, figure \ref{DensityModelMachin}, and left plate of figure \ref{DensityInversionMachin}). Notice, the contrast of densities between volcanic conduit and the surrounding rock. The left panel of this figure displays the best model obtained from the inversion. 

\begin{figure}[h!]
\begin{center} 
{\includegraphics[width=0.45\textwidth]{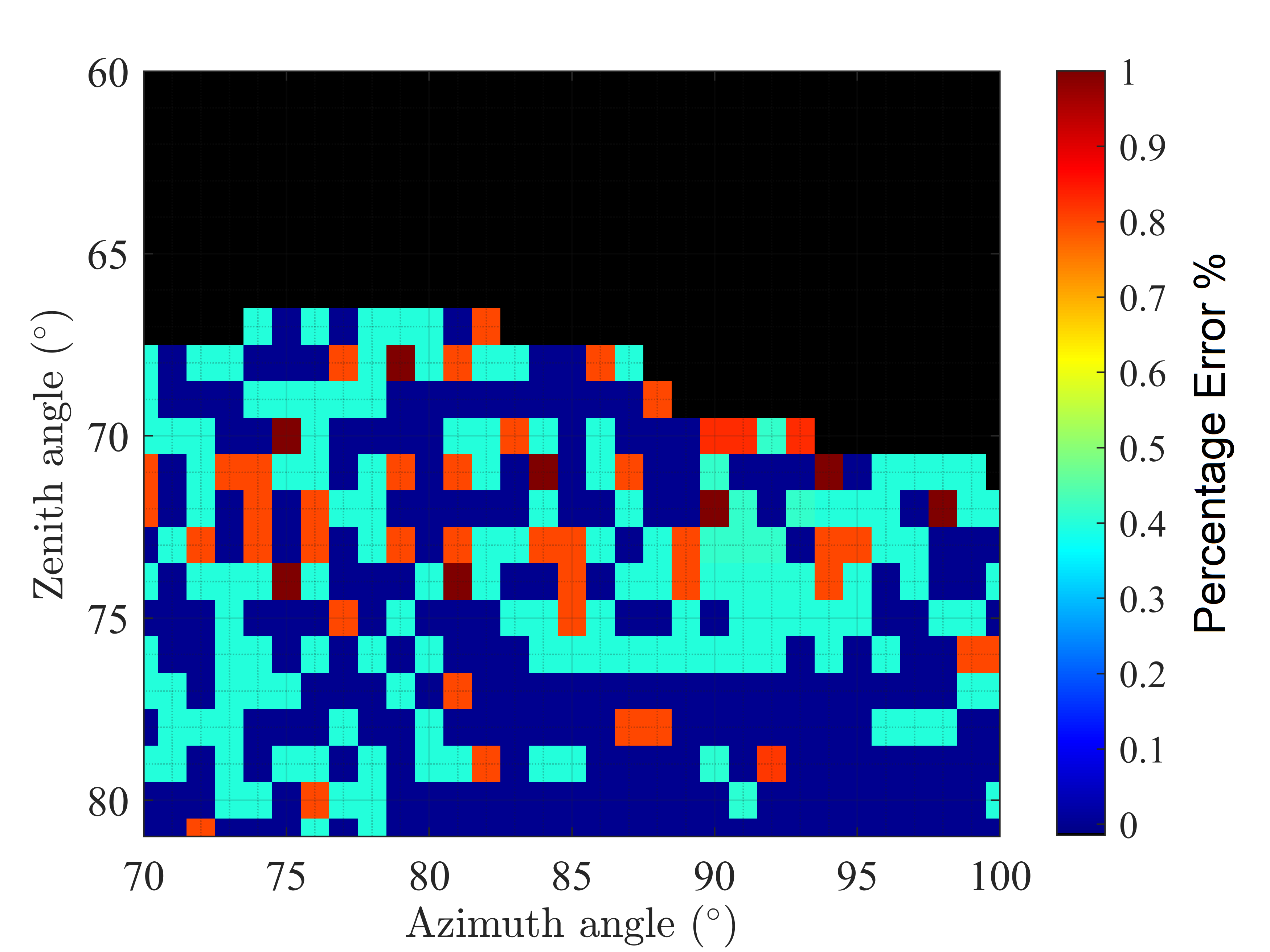}}
\caption{Error between the ``observed'' number of muons and the one predicted by the geophysical inversion. The maximum relative error was $1\%$.}
  \label{ErrorMachin}
\end{center}  
\end{figure}

Strictly, the algorithm should be implemented starting from the detected muon number at the observation point and not from a density profile.  However, because we have simulated this number of muons, we can know the precise density profile (displayed in the left plate of figure \ref{DensityInversionMachin}) and compare it with the one obtained from the inversion process. While the test scenario is synthetic, it employs a full volcano model, built from existing geological information, including the data density information collected from the field.  The resulting density profile also showed a marked contrast between the volcanic duct, the embedding rock and the fumarole zone. As shown in figure \ref{ErrorMachin}, the obtained outcome is consistent with the known ``observed'' density model, having a maximum relative error of $1$\%.

\section{Conclusions}
\label{Conclusions}
The present work details an example of the Metropolis-Simulated-Annealing algorithm for the muon radiography technique. It starts from an ``observed'' muon flux and obtains the best associated inner density distribution function inside the Cerro Mach\'in. This implementation allows the exploration of inner geological structures withing the first $190$ meters  depth from the top of the volcano. 

The estimated initial density model of the dome was obtained with GEOMODELER, adapted to the Cerro Mach\'in topography and taking into account other geological surveys \citep{inguaggiato2017hydrothermal, londono2011tomografia, londono2016evidence}. We have improved this model by including the  rock densities of samples taken from the crater, the dome and the areas associated with fumaroles.

The results of the geophysical inversion technique are essential as they serve as a guide in showing where and how the boundaries of the 3D geologic model may need to be adjusted, so as to address density excesses or deficiencies. In order to increase the geological relevance of geophysical models, we incorporate meaningful geological data into the modelling process to guide it toward a result consistent with the observed geology. One approach that we employed in this study was to use the 3D geologic model as a constraint on the geophysical modelling. The geologic model was generated using multiple geoscience datasets, including geological and structural mapping; interpreted geological sections based on 1:25.000 scale geological maps, and field samples carried out for the project. 

The inversion algorithm correctly reconstructed the density differences inside the Mach\'in, within a $1\%$ error concerning our initial simulation model, giving a remarkable density contrast between the volcanic duct, the encasing rock and the fumaroles area. We can obtain this error value because we use a semi-empirical model for impinging atmospheric muons, followed by a calculation of their energy loss passing through the geologic structure. We found the minimum energy needed to cross a pre-defined thickness of rock and estimated the muon flux emerging from the volcanic building. 

Future works will focus on improving the spatial coverage to increase the model's resolution. Furthermore, given the large non-uniqueness of the problem and the limited number of sites for muography in Cerro Mach\'in, we only interpret our model in terms of low- and high-density contrasts. We will also concentrate on the joint inversion with data from other geophysical techniques (i.e. gravimetry, electrical resistivity tomography, among others) which will have the ability to resolve densities depending on the network coverage, with a resolving power rapidly decreasing with depth.
\section*{Acknowledgments}
The authors acknowledge  financial support of  Departamento Administrativo de Ciencia, Tecnolog\'{\i}a e Innovaci\'on of Colombia (ColCiencias) under contract FP44842-082-2015 and to the Programa de Cooperaci\'on Nivel II (PCB-II) MINCYT-CONICET-COLCIENCIAS 2015, under project CO/15/02.  We are also very thankful to LAGO and to the Pierre Auger Collaboration for their continuous support.  The simulations in this work were partially possible thanks to The Red Iberoamericana de Computaci\'on de Altas Prestaciones (RICAP, 517RT0529), co-funded by the Programa Iberoamericano de Ciencia y Tecnolog\'{\i}a para el Desarrollo (CYTED) under its Thematic Networks Call. We acknowledge financial support from STFC (UK), grants ST/R002606/1. We also thank the computational support from the Universidad Industrial de Santander (SC3UIS) High Performance and Scientific Computing Centre. Finally, we would like to thank Vicerrector\'{\i}a Investigaci\'on y Extensi\'on Universidad Industrial de Santander for its permanent sponsorship. DSP would like to thank the School of Physics, the Grupo de Investigaci\'on en Relatividad y Gravitaci\'on, Grupo Halley and Vicerrector\'{\i}a Investigaci\'on y Extensi\'on of the Universidad Industrial de Santander for the support and hospitality during a post-doctoral fellowship.

\appendix
\section*{Appendix A}
\begin{table}[H]
    \centering
    \caption{{\large {\bf Forward modelling pseudo-code}}}
    \begin{tabular}{l}
    \texttt{input} \\
    $\Phi(\theta, p);$ \texttt{open sky muon flux}  \\
    $L$ \texttt{distance travelled within the volcano} \\ 
    $\theta$ and $\phi;$ \texttt{Zenith and Azimuth angle}\\ 
    \texttt{Recording time} $\Delta t$ \\
    \texttt{Begin} \\
    $\bar{\rho_0};$ \texttt{Start with an average inner density}  \\ 
    $\varrho_0;$ \texttt{Calculate the opacity}   \\
    $E_{min};$ \texttt{Calculate the minimun energy needed to cross} $L$ \\
    $I;$ \texttt{Calculate the integrated muon flux}  \\
    $\mathcal{T};$ \texttt{Calculate the aceptance based on the muon trajectories} \\
    $N(\varrho);$ \texttt{Calculate the number of impinging muons} \\
    \texttt{End}
    \end{tabular}
    \label{ForwardModPseudo}
\end{table}

\section*{Appendix B}
\begin{table}[H]
    \centering
    \caption{{\large {\bf Inverse problem pseudo-code}}}
    \begin{tabular}{l}
     \\   
    \texttt{Input from forward modelling} \\
    $N_{obs}$ \texttt{number of observed muons} \\ \hline 
    \texttt{Begin} \\
    $T \leftarrow T_{initial};$ \texttt{Set the initial evolution parameter} $T$  \\
    $r \leftarrow RAM[0,1];$ \texttt{Generate random number}  \\
    $\bar{\rho} ;$  \texttt{Select the initial simulated density model}  \\ 
    $N_{sim}(\bar{\rho});$ \texttt{Estimate the initial number of muons} \\
    $E(\bar{\rho}) = \left\| N_{obs} -N_{sim}(\bar{\rho}) \right\|;$ \texttt{Calculate the initial cost function} \\    
    \texttt{LOOP for}  $T$ \texttt{from} $T_{initial}$ to $T_{final}$ \texttt{with and step} $\Delta T_{k} \, ;$ \texttt{The ``cooling'' process} \\ 
    \, \texttt{LOOP for}  $I$ \texttt{from} $I_{initial}$ to $I_{final}$ \texttt{with step} $\Delta I \, ;$ \texttt{The refinement proces} \\ 
        \quad $r \leftarrow RAM[0,1];$ \texttt{Generate a random number}  \\
        \quad $\bar{\rho}_{r}(T);$ \texttt{Select a neighbour simulated density model}  \\ 
        \quad  $N_{sim}(\bar{\rho}_{r}, T);$ \texttt{Estimate the number of muons for a random neighbour model } \\
        \quad  $E(\bar{\rho}_{r},T) = \left\| N_{obs} -N_{sim}(\bar{\rho}_{r}, T) \right\|;$ \texttt{Calculate the cost of a random neighbour model} \\
        \quad $\Delta E(\bar{\rho}_{r}, T) = E(\bar{\rho}, T) -E(\bar{\rho}_{r},T);$ \texttt{Calculate the random model energy difference} \\
        \quad  \texttt{If} $\Delta E(\bar{\rho}_{r}, T) \leq 0$  \\
        \quad \texttt{then} \\
            \quad \quad $\bar{\rho}(T) \leftarrow \bar{\rho}_{r}(T);$  \texttt{Set as a better density value} \\
            \quad \quad $N_{sim}(\bar{\rho}, T) \leftarrow N_{sim}(\bar{\rho}_{r}, T);$ \texttt{Set as better number of muons value}\\
            \quad \quad $E(\bar{\rho},T) \leftarrow E(\bar{\rho}_{r},T);$ \texttt{Set the better cost function value} \\  
        \quad \texttt{else} \\ 
            \quad \quad \texttt{Calculate} $P= \mathrm{exp}\left( -\frac{\Delta E(\bar{\rho}_{r}, T)}{ T} \right)$ \texttt{the probability for the model admission} \\
            \quad \quad \texttt{Generate} $r_{discrim} \leftarrow RAM[0,1]$ \texttt{random number}  \\
            \quad \quad \texttt{If} $P > r_{discrim} $ \\
            \quad \quad \texttt{then} \\
            \quad \quad \quad $\bar{\rho}(T) \leftarrow \bar{\rho}_{r}(T)$   \\
            \quad \quad  \quad $N_{sim}(\bar{\rho}, T) \leftarrow N_{sim}(\bar{\rho}_{r}, T)$\\
            \quad \quad  \quad $E(\bar{\rho},T) \leftarrow E(\bar{\rho}_{r},T)$  \\   
        \quad \quad \texttt{EndIf} \\
    \quad  \texttt{EndIf} \\
    \, $I \leftarrow I - \Delta I$ \\
    \, \texttt{EndLOOP} $I$ \\
    $T \leftarrow T + \Delta T_{k}$ \texttt{decrease temperature after several iterations} \\
    \texttt{EndLOOP} $T$ \\
    \texttt{End}
    \end{tabular}
    \label{InverseModPseudo}
\end{table}

\section*{Appendix C}
\begin{table}[H]
    \centering
    \caption{{\large {\bf $T_{initial}$ pseudo-code}}}
    \begin{tabular}{l}
    \texttt{Begin} \\
    $T \leftarrow T_{initial};$ \texttt{Set a particular value for the evolution parameter} $T$  \\
    $r \leftarrow RAM[0,1];$ \texttt{Generatea first random number}  \\
    $\bar{\rho} ;$  \texttt{Select a first density model}  \\ 
    $N_{sim}(\bar{\rho});$ \texttt{Estimate the number of muons} \\
    $E(\bar{\rho}) = \left\| N_{obs} -N_{sim}(\bar{\rho}) \right\|;$ \texttt{Calculate the first cost function} \\    
    $r \leftarrow RAM[0,1];$ \texttt{Generate a second random number}  \\
    $\bar{\rho}_{r}(T);$ \texttt{Select a neighbour density model}  \\ 
    $N_{sim}(\bar{\rho}_{r}, T);$ \texttt{Estimate the number of muons for the neighbour model } \\
    $E(\bar{\rho}_{r},T) = \left\| N_{obs} -N_{sim}(\bar{\rho}_{r}, T) \right\|;$ \texttt{Calculate the cost of the neighbour model} \\
    $\Delta E(\bar{\rho}_{r}, T) = E(\bar{\rho}, T) -E(\bar{\rho}_{r},T);$ \texttt{Calculate the random model energy difference} \\
        \quad  \texttt{If} $\Delta E(\bar{\rho}_{r}, T) \leq 0$  \\
        \quad \texttt{then} \\
            \quad \quad $\bar{\rho}(T) \leftarrow \bar{\rho}_{r}(T);$  \texttt{Set as a better density value} \\
            \quad \quad $N_{sim}(\bar{\rho}, T) \leftarrow N_{sim}(\bar{\rho}_{r}, T);$ \texttt{Set as better number of muons value}\\
            \quad \quad $E(\bar{\rho},T) \leftarrow E(\bar{\rho}_{r},T);$ \texttt{Set the better cost function value} \\  
        \quad \texttt{else} \\ 
            \quad \quad \texttt{Calculate} $P= \mathrm{exp}\left( -\frac{\Delta E(\bar{\rho}_{r}, T)}{ T} \right)$ \texttt{the probability for the model admission} \\
            \quad \quad \texttt{Generate} $r_{discrim} \leftarrow RAM[0,1]$ \texttt{random number}  \\
            \quad \quad \texttt{If} $P > r_{discrim} $ \\
            \quad \quad \texttt{then} \\
            \quad \quad \quad $\bar{\rho}(T) \leftarrow \bar{\rho}_{r}(T)$   \\
            \quad \quad  \quad $N_{sim}(\bar{\rho}, T) \leftarrow N_{sim}(\bar{\rho}_{r}, T)$\\
            \quad \quad  \quad $E(\bar{\rho},T) \leftarrow E(\bar{\rho}_{r},T)$  \\   
        \quad \quad \texttt{EndIf} \\
    \quad  \texttt{EndIf} \\
    \texttt{End}    
    \end{tabular}
    \label{InitialT}
\end{table}


\bibliographystyle{elsarticle-harv} 
\bibliography{Bibliography}
\end{document}